\documentclass[fleqn,usenatbib]{mnras}

\pdfoutput=1
\usepackage[T1]{fontenc}
\usepackage{ae,aecompl}

\usepackage{graphicx}   
\usepackage{amsmath}    
\usepackage{amssymb}    

\usepackage{ifthen}
\newboolean{isdraft}
\setboolean{isdraft}{false}

\usepackage{bbold}

\ifthenelse{\boolean{isdraft}}{\usepackage{todonotes}}{}

\graphicspath{{fig/} {.}}

\title[Single-Stream Regions]{The median density of the Universe}

\author[J. St\"ucker et al.]{
Jens St\"ucker,$^{1}$\thanks{E-mail: jstuecker@mpa-garching.mpg.de (MPA)}
Philipp Busch$^{1}$
and Simon D. M. White$^{1}$
\\
$^{1}$Max-Planck-Institut f\"ur Astrophysik, Postfach 1317, D-85741 Garching, Germany
}

\date{Accepted XXX. Received YYY; in original form ZZZ}

\pubyear{2017}

\begin{document}
\label{firstpage}
\pagerange{\pageref{firstpage}--\pageref{lastpage}}
\maketitle

\begin{abstract}
Despite the fact that the mean matter density of the universe has been measured to an accuracy of a few percent within the standard $\Lambda$CDM paradigm, its median density is not known even to order of magnitude. Typical points lie in low-density regions and are not part of a collapsed structure of any scale. Locally, the dark matter distribution is then simply a stretched version of that in the early universe. In this single-stream regime, the distribution of unsmoothed density is sensitive to the initial power spectrum on all scales, in particular on very small scales, and hence to the nature of the dark matter. It cannot be estimated reliably using conventional cosmological simulations because of the enormous dynamic range involved, but a suitable excursion set procedure can be used instead. For the Planck cosmological parameters, a 100 GeV WIMP, corresponding to a free-streaming mass $\sim 10^{-6}M_\odot$, results in a median density of $\sim 4\times 10^{-3}$ in units of the mean density, whereas a 10 $\mu$eV axion with free-streaming mass $\sim 10^{-12}M_\odot$ gives $\sim 3\times 10^{-3}$, and Warm Dark Matter with a (thermal relic) mass of 1 keV gives $\sim 8\times 10^{-2}$. In CDM (but not in WDM) universes, single-stream regions are predicted to be topologically isolated by the excursion set formalism. A test by direct N-Body simulations seems to confirm this prediction, although it is still subject to finite size and resolution effects. Unfortunately, it is unlikely that any of these properties is observable and so suitable for constraining the properties of dark matter.
\end{abstract}




\section{Introduction}
Let us start this article with a simple thought experiment: Imagine we were able to reliably measure the mass density in small volume elements, let us say cubes with a side length of a kilometre. And let us further assume we would be able to do this measurement anywhere in the universe. Now if we would do this measurement for a large number of randomly placed cubes, what would the distribution of their densities look like? Or rephrasing: \emph{What is the one-point density distribution of the Universe at very high resolution?} What is its median density? What is its shape? What determines the behaviour of its high- and low-density tails? And if we knew the distribution, could we learn something about dark matter?

Although such a measurement is not possible today nor will it be possible anywhere in the near future, trying to answer this question from purely theoretical arguments turns out to give considerable insight. It helps paint a simple  picture of what is happening in the majority of the volume - from the largest to the smallest scales. We here propose an excursion set formalism which provides such a qualitative picture and further enables us to estimate the unsmoothed density distribution of the universe.

In the current best fitting model of cosmic structure formation, the main gravitating component is dark matter. In the early universe it is distributed almost homogeneously with only small perturbations from the mean density. At this time the distribution of density perturbations is expected to be given by a simple Gaussian distribution.

\begin{figure}
	\includegraphics[width=\columnwidth]{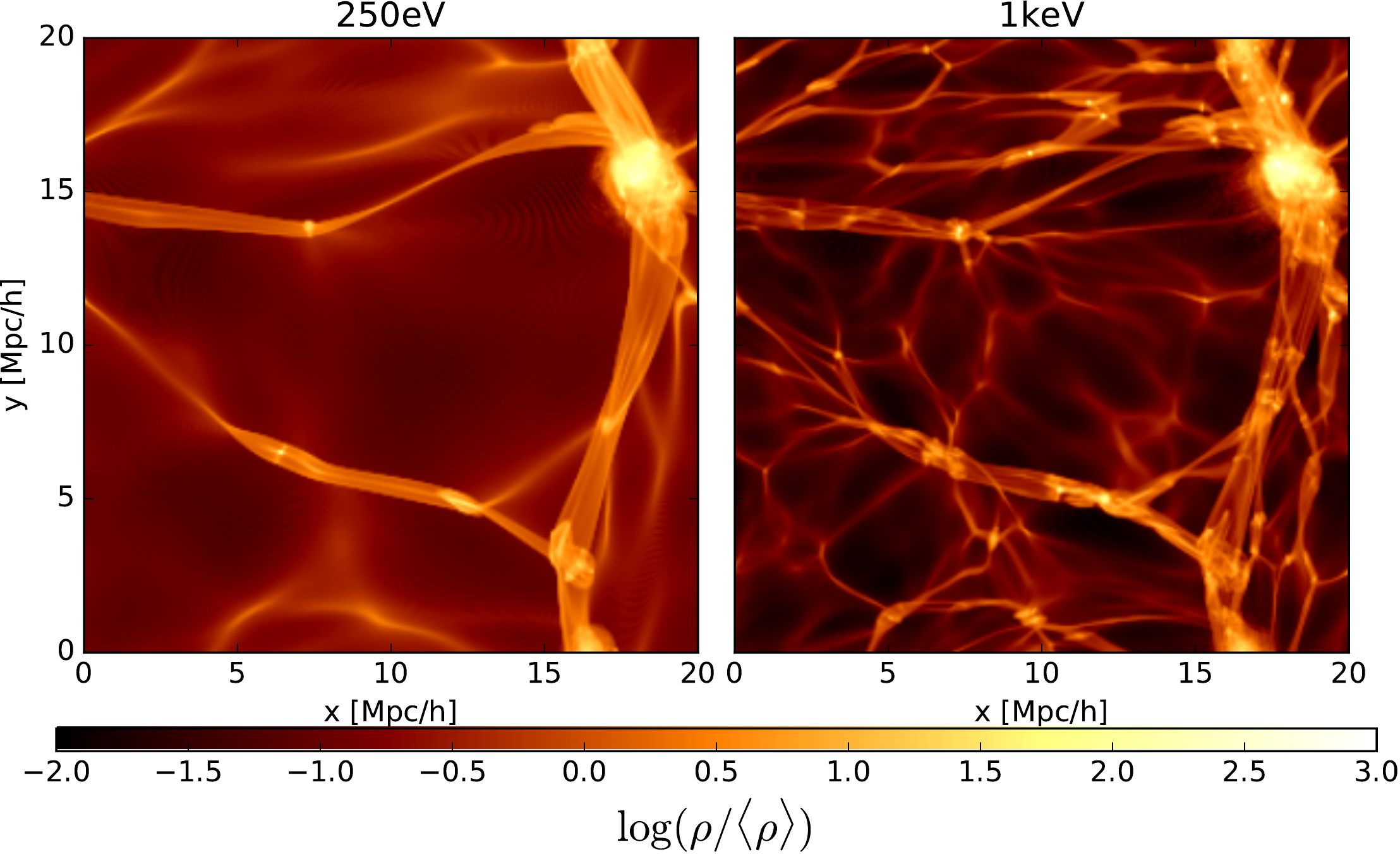}
    \caption{A razor thin slice through two WDM simulations with different thermal cut-offs in the power spectra corresponding to a 250eV thermal relic (left) and a 1keV thermal relic (right).  In the un-collapsed single-stream regions the density distribution depends strongly on the dark matter model and its Lagrangian smoothing scale.}
    \label{fig:wdm_slice}
\end{figure}

However as the universe expands, the perturbations grow - early on in a simple linear way, and thereafter in a non-linear and partially chaotic manner. While the linear regime can be well described by analytic methods, the investigation of the dark matter distribution in the non-linear regime usually requires N-Body simulations which explicitly follow the evolution of a large set of tracer particles in a three dimensional cosmological volume. 

The power spectrum of density perturbations is extremely flat for cold dark matter cosmologies in the sense that density perturbations on all scales from hundreds of megaparsecs down to a thermal smoothing scale of e.g. parsecs (for WIMPs) are relevant to determine what happens to the unsmoothed density field in the non-linear regime. Therefore a cosmological simulation to follow the unsmoothed density field of WIMP-like dark matter, would need to resolve about 8 orders of magnitude in spatial scale, requiring of order $10^{24}$ resolution elements which is still far from what is possible.

This problem is usually tackled by smoothing the initial conditions of dark matter simulations on a relatively large length scale - either explicitly by introducing a cut-off scale into the power spectrum as in warm dark matter simulations, or implicitly by the Nyquist frequency of the mesh that samples the initial density field. The conclusions that can be made from these simulations are then limited to features that do not depend on initial perturbations that are smaller than this Lagrangian smoothing scale. As most observations involve a relatively large smoothing anyway, the smoothing in Lagrangian space is usually of little importance for the comparison with observations. However, in our thought experiment we are asking explicitly for the unsmoothed density field.

The unsmoothed density field depends strongly on the small scale cut-off of the dark matter power spectrum (which is equivalent to a Lagrangian smoothing scale). To illustrate this we show in Figure \ref{fig:wdm_slice} a thin slice through two warm dark matter simulations with different free-streaming scales. The smaller the smoothing scale, the more diffuse material fragments into small scale structures, and the lower the typical density of the universe becomes.

\begin{figure}
	\includegraphics[width=\columnwidth]{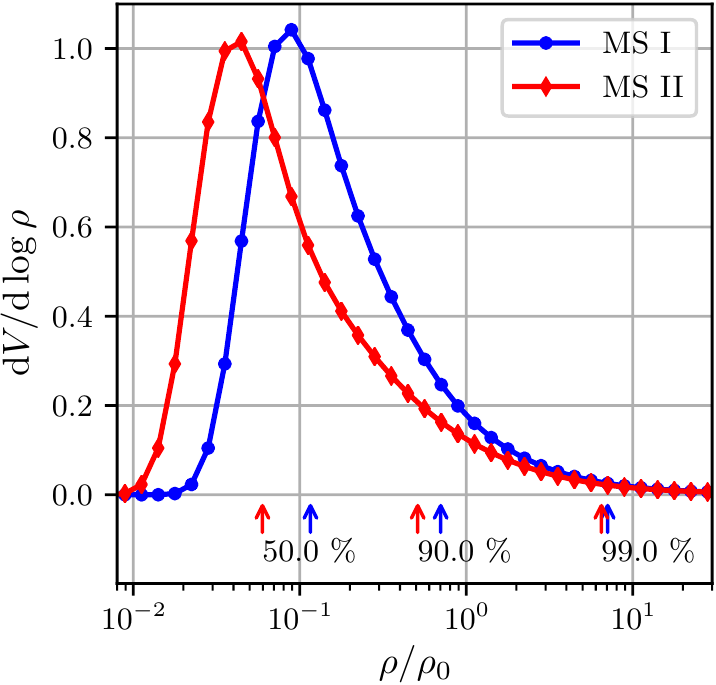}
    \caption{Comparison between the volume-weighted density distribution of the particles in the Millennium I (MS I) and Millennium II  (MS II) simulations, using the volume of each particle's Voronoi cell to calculate its density. The density distribution in classical cold dark matter simulations is still far from from converged with resolution.}
    \label{fig:msi_msii}
\end{figure}

While the Lagrangian smoothing scale is incorporated explicitly in these two simulations, it is also present implicitly in all classical cold dark matter simulations. If the resolution of a cold dark matter simulation is changed, the maximum spatial frequency of the imposed initial perturbations shifts, leading to additional small scale structure which strongly modifies the density distribution. As an example of this we show the volume-weighted density distribution of the Millennium Simulation \citep{springel_simulating_2005} in comparison to the much higher resolution Millennium Simulation II \citep{boylan-kolchin_resolving_2009} in Figure \ref{fig:msi_msii}. Here the density field is approximated by attributing the volume of each cell in a Voronoi tessellation of the particle distribution to the particle at its centre and using this to provide a density estimate. Additionally we show the 50-, 90- and 99-percentiles of the distributions. It is evident that these density distributions are far from converged. While the relatively good convergence in the high density tails of these distributions was previously discussed by \citet{pandey_2013}, here we show that there is a factor of 2 between the median densities. A similar factor lies between the peaks in the density distributions and the minimum particle densities ($\left(\rho_{\mathrm{min}}/\rho_{0}\right)_{MSI}=1.1\cdot 10^{-2}$ and $\left(\rho_{\mathrm{min}}/\rho_{0}\right)_{MSII}=5.6\cdot 10^{-3}$). 

It is worth noting that while most of the mass is part of collapsed structures, most of the volume is part of single-stream regions. Therefore the density distribution of the universe is mostly given by the density distribution of single-stream regions. Only the high density tail will be affected by collapsed structures. Single-stream regions are expected to be mathematically simpler, and we will focus in the remainder of this paper on predicting their density distribution.

We loosely refer to single-stream regions here as regions which have not undergone any collapse, specifically, where the Lagrangian patch has not passed through any caustic, thereby speaking of the diffuse three dimensional material between structures. This is similar to the idea of a void. The term void is, however, often used to refer to the largest under-densities in the universe after smoothing on Mpc scale or larger (e.g. \citet{Weygaert_Platen_2011}). Such voids actually contain many collapsed objects of smaller scale. In contrast our subject of interest here is to describe the regions of the universe which contain no collapsed object of any scale, motivating our definition of single-stream regions. Tracing of the detailed structure of the dark matter phase sheet has recently become possible \citep{shandarin_2012, abel_2012} allowing the stream multiplicity to be measured in simulations, and giving an interesting new perspective on structures in the universe \citep{ramachandra_shandarin_2017}.

We propose an excursion set formalism that allows prediction of the density distribution of single-stream regions. The formalism defines a collapse criterion which detects the first caustic crossing of a particle which occurs as it first becomes part of a two dimensional sheet-like structure (a pancake). It then checks whether this criterion is fulfilled by the smoothed linear density field at the point corresponding to a particular particle for \emph{any} Lagrangian smoothing scale. If it is not fulfilled on any scale, the particle is assumed to be part of a single-stream region. In that case we expect the local density to be well described by a simple model such as the Zeldovich approximation or the triaxial collapse model which we introduce here. We use this excursion set formalism to predict the density distribution of single-stream regions, and the total amount of mass expected within single-stream regions. 

Another interesting question that can be answered within this context is whether single-stream regions form distinct regions enclosed by collapsed structures (i.e. by multi-stream structures), or whether they form one connected infinitely large percolating  region. This subject has already been investigated by \citet{falck_neyrinck_2015} who find that the single-stream regions in their simulations percolate. In contrast we show in section \ref{sec:percolation} that the excursion set formalism predicts that single-stream regions do not percolate in cold dark matter universes. The regime where single-stream regions stop percolating lies beyond the resolution limit of the simulations of \citet{falck_neyrinck_2015}. We attempt to test this regime with an N-Body simulation. When we test for percolation in Eulerian space, we find that the sizes of individual single-stream regions depend significantly on resolution parameters. However, a percolation test in Lagrangian space, which we consider more robust against numerical artefacts, shows no percolation. We thus infer that single-stream regions do not percolate in the continuum limit of cold dark matter.

\section{An Excursion Set Formalism for Single-Stream Regions}
In this section we introduce an excursion set formalism that can be used to predict the density distribution of single-stream regions. We briefly review classic excursion set formalisms, then introduce two alternative models for the collapse barrier, and summarize the mathematical background needed for the six dimensional random walk of the deformation tensor. Finally, we note that the predictions of the excursion set models depend only on the variance of the unsmoothed linear dark matter density field.
\subsection{Excursion Set Formalisms}
Excursion Set formalisms have long been used as simplified models of structure formation in the non-linear regime. Probably the most prominent one is the extended Press-Schechter formalism, hereafter EPS, by \citet{bond_1991} which has been used to predict conditional and unconditional halo mass functions and halo clustering bias. The unconditional mass function turns out to be very similar to that originally obtained by \citet{press_schechter_1974}.

The EPS formalism is based on a simple assumption derived from the spherical collapse model: any uniform spherical perturbation which has a linear-theory density contrast larger than $\delta_c = 1.68$ is assumed to have collapsed. Given a realisation of a linear cosmological density field all particles for which $\delta > \delta_c$ for some smoothing scale are assumed to be part of a halo in a fully non-linear calculation. The mass of the halo is assumed to correspond to the largest smoothing scale for which the collapse criterion is satisfied. 

In the simplest version of the EPS formalism the largest scale is identified by smoothing with a top-hat filter in Fourier space. One starts with an infinitely large smoothing length scale $R_s \propto k_s^{-1}$ and decreases it smoothly. The density contrast at a single point then effectively makes an uncorrelated random walk. If the density contrast first crosses the barrier $\delta_c$ the particle is assumed to be part of a halo with Lagrangian size corresponding to the smoothing scale $R_s$ at first crossing. In the EPS model the distribution of $R_s$ at first crossing thus determines the halo mass function.

Despite its simplicity the excursion set formalism has been shown to give reasonably good predictions for the halo mass function \citep[e.g.][]{zentner_2007}. It has been improved by introducing more sophisticated barriers which consider deviations from spherical symmetry \citep{sheth_tormen_2001}. Arguably a major reason why it works so well is the way large scale modes interact with small scale modes in the non-linear regime. Large modes strongly influence what is happening on small scales, while small scales barely influence any large scale structure. Therefore if something can be determined to collapse when the density field is smoothed on a large scale, it will almost certainly also collapse in the unsmoothed density field, since the smaller scale perturbations do not influence the large scale structure. Qualitatively this can be seen in Figure \ref{fig:wdm_slice} where the dominant large structures remain almost unchanged by decreasing the Lagrangian smoothing scale.

While the EPS formalism seems to describe the formation of haloes reasonably well, it does not try to explain the behaviour of the material outside of haloes. This material can either be part of string-like filaments, planar sheet-like ''pancakes'' or diffuse three dimensional single-stream regions. Single-stream regions are regions which have not yet collapsed on any scale.

We propose an excursion set formalism here that tries to predict properties of these single-stream regions. It considers a particle to be part of a single-stream region if it does not fulfil a collapse criterion on any length scale. In this case the collapse criterion does not mark the point where a particle becomes part of a halo, but it marks the point where it goes through its first caustic - which normally happens in a pancake or filament. We consider two models for the collapse criterion here (1) the Zeldovich approximation and (2) a triaxial collapse model. Further we assume the stream density of particles to be given by these simple models if they are part of a single-stream region. This allows us to evaluate different statistics of single-stream regions. 

Note that while the EPS formalism only requires following the random walk of the density, our formalism requires following the three eigenvalues of the deformation tensor. The idea of following the eigenvalues of the deformation tensor in the excursion set formalism has already been explored in previous work in the context of halo formation and its relation to the cosmic web\citep{chiueh_2001, sandvik_why_2007}. Here we will use it to learn more about single-stream regions.

\subsection{The Zeldovich Approximation}
\begin{figure}
	\includegraphics[width=\columnwidth]{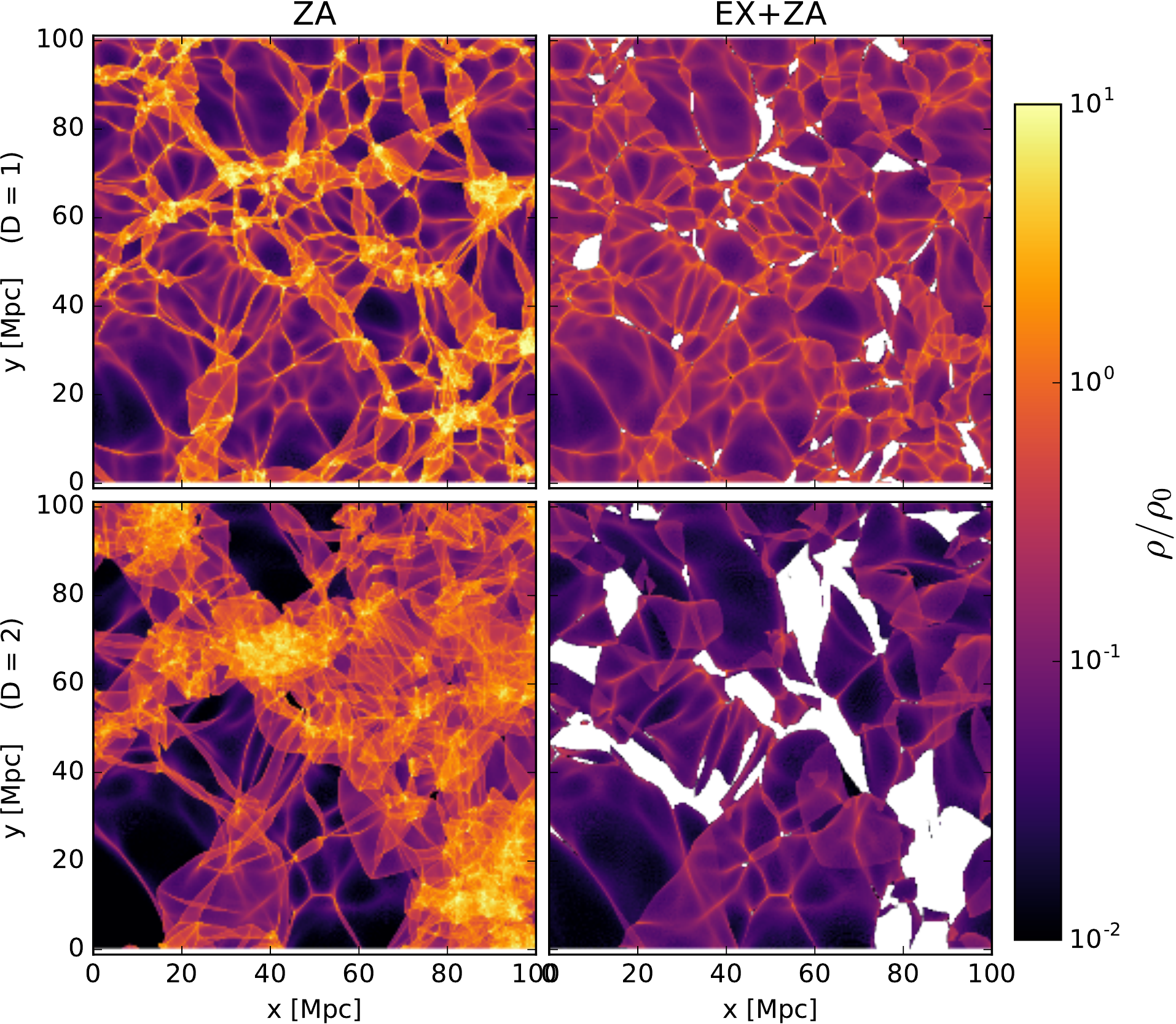}
    \caption{Left column: Evolution under the Zeldovich approximation for a two dimensional density field using a power spectrum with normalization $\sigma = 2.36$ (as defined in \eqref{eqn:sigma}) at a growth factor of $D=1$ (top) and $D=2$ (bottom). Right: the same original density field, but cutting out particles that were classified as belonging to a multistream region with our excursion set formalism. The linearly evolved two point correlation function in this two dimensional test problem corresponds to that of a $m_X = 250eV$ WDM cosmology. }
    \label{fig:zeldovich_slice}
\end{figure}
As a first idea to model single-stream regions we consider the Zeldovich approximation. The Zeldovich approximation relates the comoving Lagrangian coordinates $\vec{q}$ at an initial time ($a = 0$) to the comoving Eulerian coordinates $\vec{x}$ at a later time
\begin{align}
  \vec{x}(a) = \vec{q} + D(a) \vec{s}(\vec{q})
\end{align}
where $D(a)$ is the linear growth factor normalized to $1$ at the present day $D(a = 1) = 1$ and $\vec{s}(\vec{q})$ is the displacement field which is approximated within the Zeldovich approximation by the gradient of a potential field
\begin{align}
  \vec{s}(\vec{q}) &= - \vec{\nabla_q} \phi (\vec{q})
\end{align}
where $\phi$ is the comoving displacement potential which is assumed to be an homogeneous isotropic Gaussian random field. These expressions are exact in linear theory. The comoving densities at any scale factor $a$ can be evaluated as
\begin{align}
  \frac{\rho(a)}{\rho_0} &= \left|\det \left(\frac{d\vec{x}}{d\vec{q}}\right)\right|^{-1} \\
                         &= \left|\det \left(\mathbb{1} - D(a) \frac{d \vec{s}}{d\vec{q}} \right) \right|^{-1} \\
                         &= \left| (1 - D(a) \lambda_1) (1 - D(a) \lambda_2) (1 - D(a) \lambda_3) \right|^{-1} \label{eqn:deformation-density}
\end{align}
where $\lambda_1 \geq \lambda_2 \geq \lambda_3$ are the eigenvalues of the deformation tensor
\begin{align}
  d_{ij} = \frac{ds_i}{dq_j} \text{\quad .}
\end{align}
Within the Zeldovich approximation a particle passes its first caustic when $D(a) \lambda_1 = 1$. The particle then becomes part of a structure which is collapsed in one dimension, but remains extended in the two other dimensions - commonly referred to as a pancake or sheet-like structure. We use this as our first collapse criterion.

In Figure \ref{fig:zeldovich_slice} we show the excursion set + Zeldovich approximation (from now on EX+ZA) approach in practice for a two dimensional density field. Even though a large part of the density field has undergone shell-crossing, and is lacking any decelerating forces within the ZA, the ZA still appears to give a reasonable qualitative picture of regions that have been classified as single-stream regions within the EX+ZA formalism.

\subsection{Triaxial Collapse Model}
While the Zeldovich approximation gives a good qualitative description for the behaviour of single-stream regions, it fails quantitatively in the non-linear regime as we shall see in section \ref{sec:simulations}. We have therefore developed a triaxial model for the evolution of infinitesimal volume elements:
\begin{align}
  \dot{x}_i &=   a^{-2} p_i  \label{eqn:tc_1} \\
    \dot{p}_i &= -  \frac{4 \pi G}{3} \rho_{bg} a^{-1} x_i \left( \delta + \alpha(t) (3 \lambda_i - \delta_0) \right) \\
    \delta &= \frac{1}{x_1 x_2 x_3} - 1 \label{eqn:tc_3}
\end{align}
for $i = 1,2,3$ where the $x_i$ represent the individual Lagrangian to Eulerian expansion factors of the three principal axes of a volume element, and the $p_i$ are the related momentum variables. $\lambda_i$ are the eigenvalues of the deformation tensor, and $\delta_0 = \lambda_1 + \lambda_2 + \lambda_3$ is the linear density contrast at a scale factor $a=1$. $\rho_\text{bg}$ is the (time dependent) mean matter density of the universe and $\delta$ is the relative over-density of the considered volume element. We derive this model in Appendix \ref{app:triaxial_collapse}. It describes the general evolution of a single-stream volume element under the influence of external tidal forces. The time dependent factor $\alpha(t)$ parametrizes how the external tidal field grows with time. In linear theory $\alpha(t) = D(t)$. While this is certainly correct in early stages of evolution, it leads to strongly over-estimated tidal forces in the non-linear regime (compare Appendix \ref{app:tidal}). To limit the external tidal field in the non-linear regime we instead use
\begin{align}
  \alpha(t) &= \frac{D(t)}{1 + |\delta_0| D(t)} \text{\quad .} \label{eqn:fade}
\end{align}
Note that other choices for $\alpha(t)$ are possible, and they lead to similar results. It is mostly important that linear theory is recovered for early stages and that the external tidal field becomes sub-dominant in the strongly non-linear regime.

This triaxial model is conceptually very similar to the ellipsoidal collapse model in \citet{bond_myers_1996}. Both models try to follow the evolution of a volume element that is subject to a tidal field that is described by the eigenvalues of the deformation tensor. However, the BM96 model assumes that the described perturbation has an ellipsoidal shape. This assumption can be dropped when speaking of an infinitesimal volume element which results in the simpler differential equations in \eqref{eqn:tc_1} - \eqref{eqn:tc_3}. 

\begin{figure}
  \includegraphics[width=0.49\columnwidth]{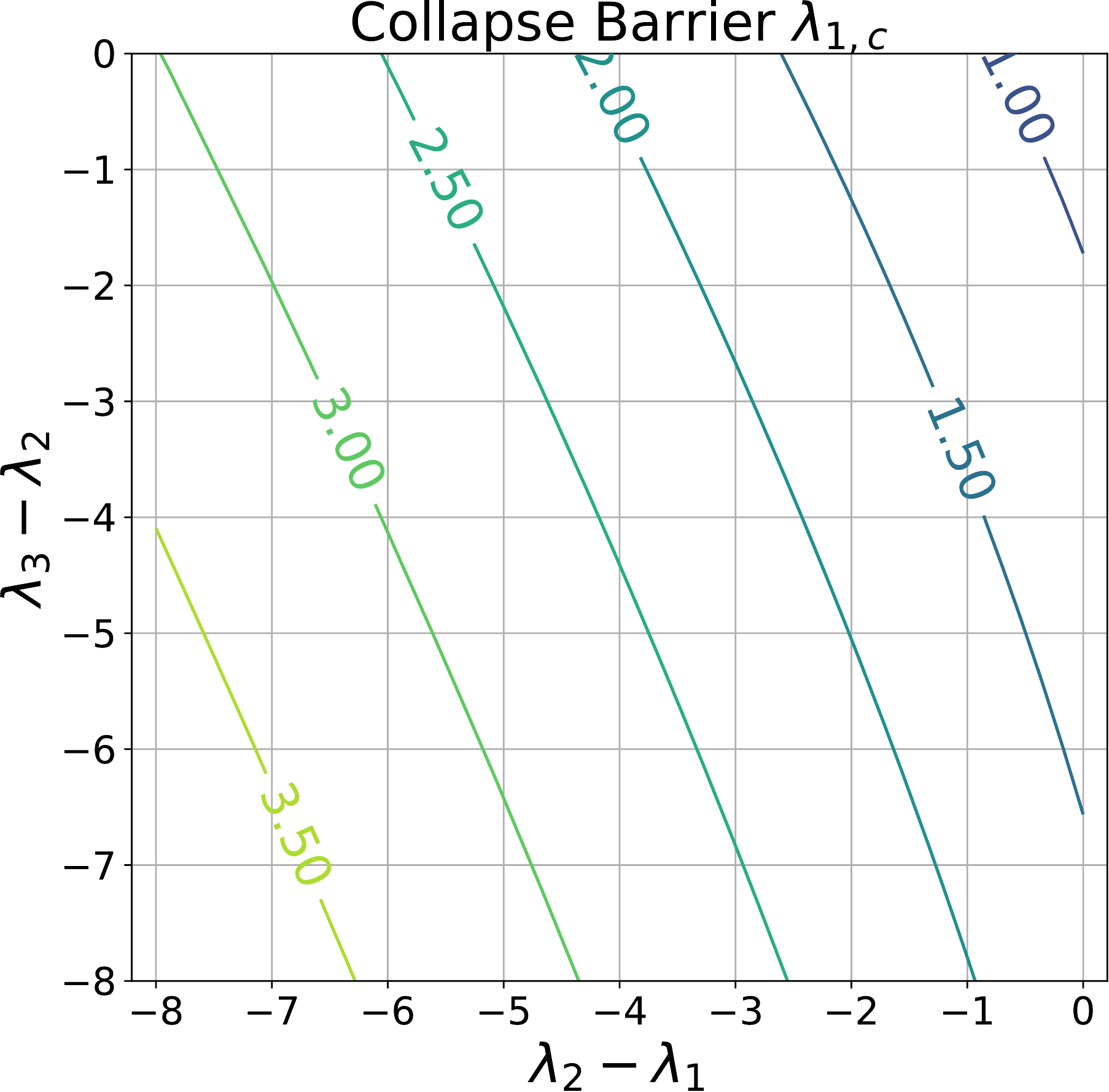}
  \includegraphics[width=0.49\columnwidth]{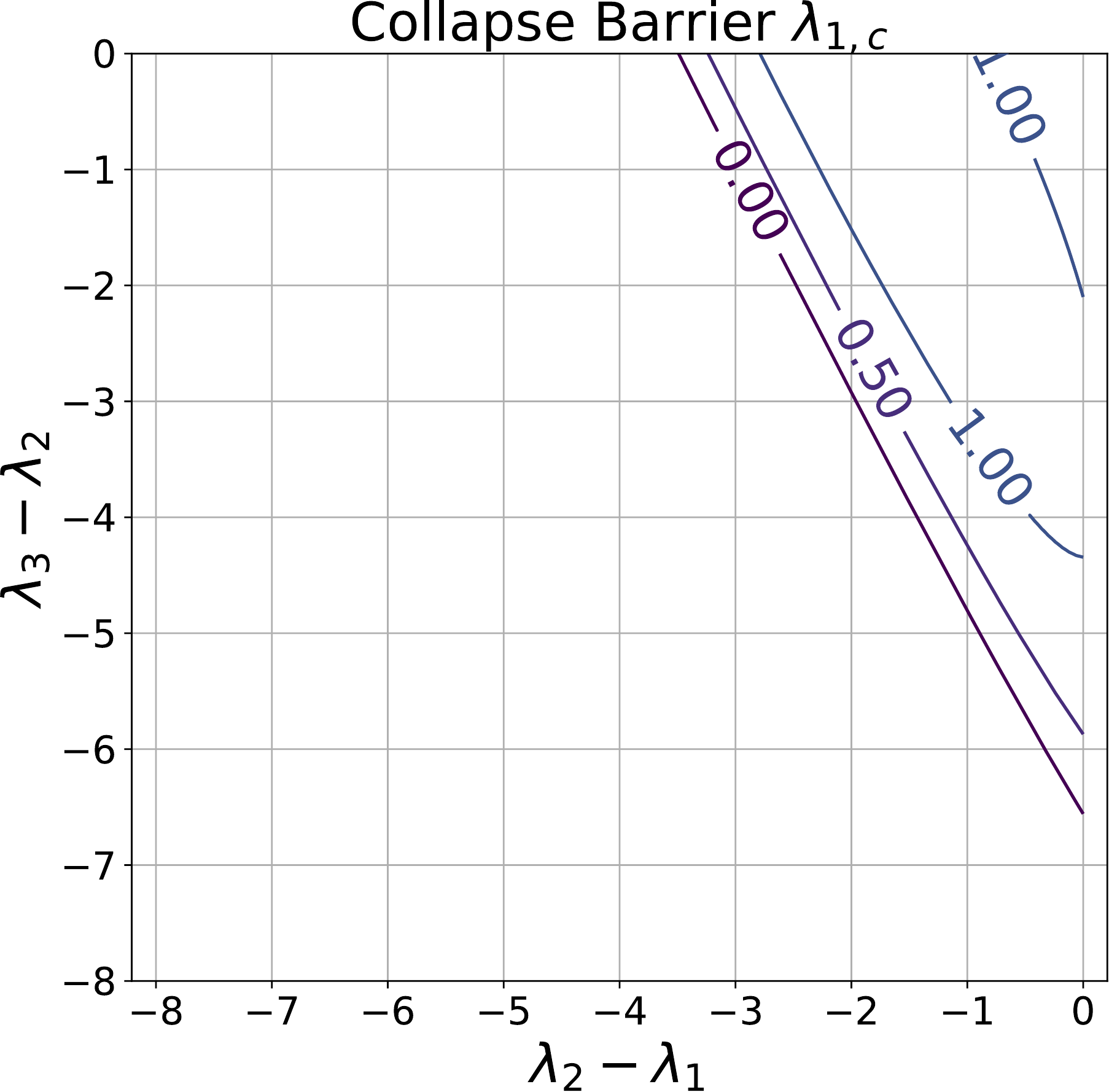}
  \caption{The collapse barrier for the biggest eigenvalue $\lambda_1$ of the deformation tensor as a function of the two smaller eigenvalues. If $\lambda_1 > \lambda_{1,c}$ a particle is predicted to be part of a collapsed structure by the triaxial collapse model. Left: with fading tidal field as in \eqref{eqn:fade}. If the two larger axes get stretched by the tidal field $\lambda_3 \leq \lambda_2 \ll 0$ the threshold $\lambda_{1,c}$ becomes larger and the collapse thereby more unlikely.  Right: Triaxial collapse with the tidal field from linear theory ($\alpha(t) = D(t)$). The tidal field becomes too large and can lead to premature collapse in cases with $\lambda_3 \leq \lambda_2 \ll 0$. Note that the spherical collapse barrier is correctly reproduced for $\lambda_3 = \lambda_2 = \lambda_{1,c} \approx 1.68/3$.}
  \label{fig:collapse_barrier}
\end{figure}

To be able to evaluate the density prediction of the triaxial collapse model as a function of the eigenvalues of the deformation tensor $\rho_s(\lambda_1, \lambda_2, \lambda_3, a=1)$ we integrate the equations of motion for a large set of parameters, store them in an interpolation table and interpolate it to the requested values later. Further we determine the collapse barrier which we define as the point where the smallest axis becomes $x_1 = 0.1$ and parametrize it as a threshold for the largest eigenvalue $\lambda_1$ depending on the two smallest eigenvalues of the deformation tensor. In Figure \ref{fig:collapse_barrier} we show this collapse barrier $\lambda_{1,c} (\lambda_2, \lambda_3)$ for the model with fading field (as defined in equation \eqref{eqn:fade}) and for the tidal field from linear theory ($\alpha(t) = D(t)$). The model that uses the pure tidal field from linear theory can apparently lead to premature collapse for cases with strongly negative eigenvalues $\lambda_3 < \lambda_2 \ll 0$. For the excursion set formalism we therefore only consider the triaxial collapse model with fading tidal field from equation \eqref{eqn:fade}.


\subsection{The Six-Dimensional Random Walk of the Deformation Tensor}

Following \citet{chiueh_2001} for a cosmological density field with rms density fluctuation $\sigma$ a random realization of the deformation tensor can be generated by drawing six independent random variables $\{y_1, ... y_6\}$ from a normal distribution with dispersion $\sigma$ and using the transformation
\begin{align}
  d_{11} &= - \frac{1}{3} \left( y_1 + \frac{3}{\sqrt{15}} y_2 + \frac{1}{\sqrt{5}} y_3 \right)\\
  d_{22} &= - \frac{1}{3} \left( y_1 - \frac{2}{\sqrt{5}} y_3 \right)\\
  d_{33} &= - \frac{1}{3} \left( y_1 - \frac{3}{\sqrt{15}} y_2 + \frac{1}{\sqrt{5}} y_3 \right)\\
  d_{12} &= d_{21} = \frac{1}{\sqrt{15}} y_4 \\
  d_{23} &= d_{32} = \frac{1}{\sqrt{15}} y_5 \\
  d_{13} &= d_{31} = \frac{1}{\sqrt{15}} y_6
\end{align}
A random walk within the deformation tensor can then be constructed by choosing $n$ intervals between $\sigma_0 = 0$ and $\sigma_n = \sigma_{max}$ and subsequently evaluating
\begin{align}
  d^{(k)}  &= d^{(k-1)} + \Delta d_k (\Delta \sigma_k^2)
\end{align}
where one starts with $d^{(0)}_{ij} = 0$ and draws each step a random $\Delta d_k$ as explained above with  a dispersion $\Delta \sigma_k^2 = \sigma_{k}^2 - \sigma_{k-1}^2$. Note that the only cosmology dependent parameter which enters this random walk is the standard deviation of the final unsmoothed density field
\begin{align}
  \sigma_{max}^2 = \frac{1}{2 \pi^2} \int_0^\infty P(k) k^2 dk \text{\quad .}  \label{eqn:sigma}
\end{align}
The value $\sigma_{max}$ is unknown and depends on the free streaming cut-off of the considered dark matter model. Therefore, a measurement of the (unsmoothed) dark matter density within single-stream regions would directly constrain $\sigma_{max}$, allowing conclusions about the nature of the dark matter particle. Unfortunately this is unlikely ever to be possible.

In each step of the random walk we diagonalize the deformation tensor to obtain its three eigenvalues and test whether a collapse criterion is fulfilled $\lambda_1 \geq \lambda_{1,c} (\lambda_2, \lambda_3)$. For each random walk trajectory we save whether it has ever been outside of the barrier, and we also store the values of the eigenvalues $\lambda_1$,$\lambda_2$ and $\lambda_3$ at the final step $\sigma_{n}$. At the end of a random walk we assume for all single-stream particles (which have never crossed the barrier) that their density is given by
\begin{align}
  \rho_{s} = \rho_m(\lambda_1, \lambda_2, \lambda_3)
\end{align} 
where $\rho_m$ is
\begin{align}
  \rho_{za}(\lambda_1, \lambda_2, \lambda_3) = \frac{\rho_0}{(1-\lambda_1)(1-\lambda_2)(1-\lambda_3)}
\end{align}
for the Zel'dovich approximation and
\begin{align}
  \rho_{tc}(\lambda_1, \lambda_2, \lambda_3) = \frac{\rho_0}{x_1 x_2 x_3}
\end{align}
for the triaxial collapse model where the $x_i$ have been evaluated via numerical integration.

\subsection{The Thermal Cutoff}
The predictions of the excursion set formalism are determined entirely by the rms density fluctuation $\sigma$ of the unsmoothed density field. The value of $\sigma$ is not known and depends both on the physics of inflation and the particle physics properties of the dark matter. We will give estimates of $\sigma$ for different dark matter models here.
\begin{figure}
	\includegraphics[width=\columnwidth]{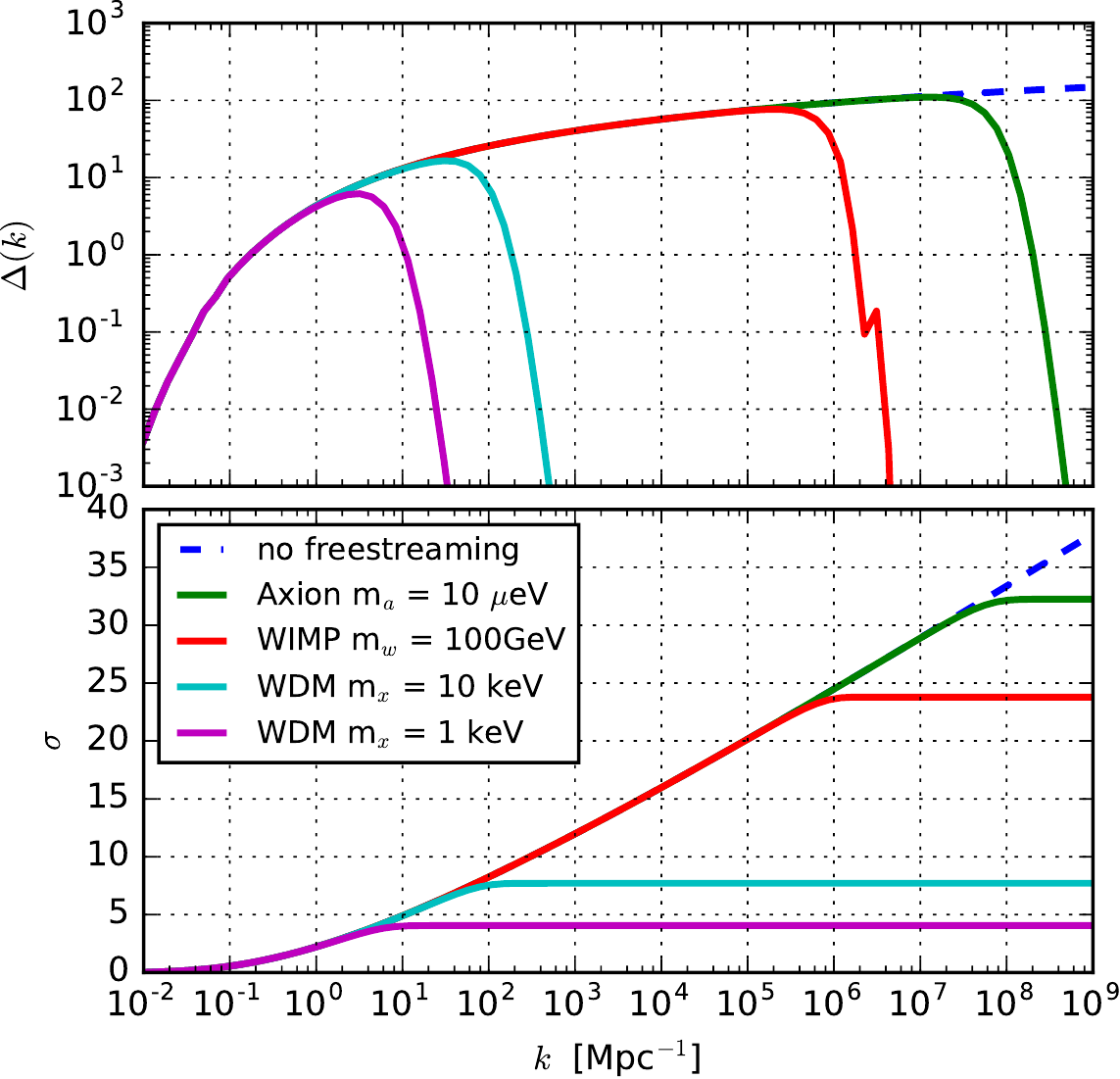}
    \caption{Top: the dimensionless linear power spectrum (top) for different different dark matter models. Bottom: the $\sigma$ values obtained by integrating the power spectra up to a scale $k$ according to \eqref{eqn:sigmak}.}
    \label{fig:darkmatter_models}
\end{figure}

Since the dark matter power spectrum has a slope close to $-3$ on small scales, the integral in \eqref{eqn:sigma} is nearly logarithmically divergent for a generic cold dark matter model that ignores the effects of free streaming (e.g. the dashed line in Figure \ref{fig:darkmatter_models}). However, most dark matter models exhibit a cut-off of the power spectrum on small scales, since small scale pertubations are smoothed out either by the effect of thermal free streaming or by quantum effects. The value of $\sigma$ will depend strongly on the scale of the cutoff. Here we consider cutoff models for warm dark matter (WDM), WIMP-based cold dark matter and axion-based cold dark matter. For the warm dark matter models we use the cold dark matter power spectrum parametrization of \citet{eisenstein_power_1999} and apply the warm dark matter thermal cutoff from \citet{bode_halo_2001}. For the WIMP models we use the \citet{eisenstein_power_1999} spectrum for $k \leq 10^2 h \text{Mpc}$ and use the small scale parametrization of \citet{green_wimpy_2005} for $k > 10^2 h \text{Mpc}$. We normalize the \citet{eisenstein_power_1999} spectrum to the cosmological value of $\sigma_8$ and choose the normalization of the \citet{green_wimpy_2005} to match $P_{e,h}(k = 10^2 h \text{Mpc}) = P_{g}(k = 10^2 h \text{Mpc})$. For the axion model we use the same cut-off parametrization as in the WDM models, with the effectively rescaled mass relation from \citet{marsh_2016} equation (118). We present the dimensionless power spectra 
\begin{align}
  \Delta(k) &= P(k)\frac{k^3}{2\pi^2} 
\end{align}
and the values of 
\begin{align}
  \sigma(k) &= \left( \int_0^{\ln(k)} \Delta(k') d\ln k'  \right)^{1/2} \label{eqn:sigmak}
\end{align}
for four different dark matter models in Figure \ref{fig:darkmatter_models}. This leads us to the total rms density fluctuations $\sigma$ which are listed in Table \ref{tab:model_data}. Note that these models are just intended to give a rough impression of the range of possible values for $\sigma$ which is quite large given the weak current constraints on the non-gravitational properties of dark matter.
\section{Test on Simulations}
\label{sec:simulations}
\begin{figure}
  \includegraphics[width=\columnwidth]{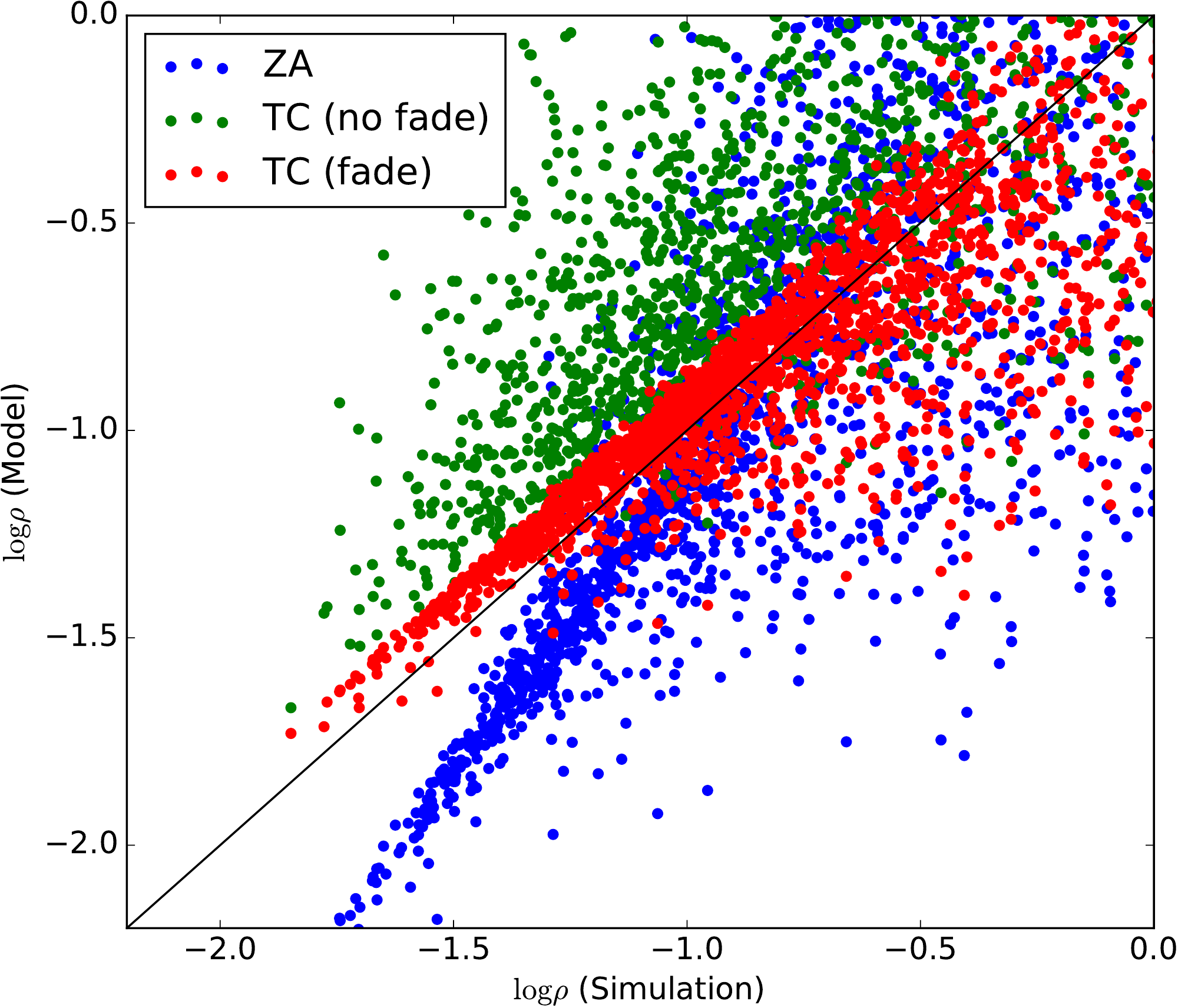}
  \caption{Predicted versus measured stream densities for particles that have not crossed a caustic at $a=1$ for different models. In blue: the Zeldovich approximation which shows a tight correlation, but systemically underpredicts the stream densities. In green: the triaxial collapse model with the tidal field from linear theory. It has a much larger scatter and seems to over-predict the densities. In red: the triaxial collapse model with fading tidal field as in \eqref{eqn:fade}. This results in a tight correlation with the right slope, but a small offset (over-prediction) in the densities. }
  \label{fig:density_correlation}
\end{figure}
To evaluate quantitatively whether the excursion set formalism produces reasonable results, we compare it with a $1$keV warm dark matter simulation with $512^3$ particles in a box with a side-length of $L=20 \text{Mpc}$. The simulation is carried out using a modified version of Gadget 3 \citep{gadget2} using the HA16 scheme \citep{hahn_angulo_2016} without refinement. We track the distortion tensor $\frac{d\vec{x}}{d\vec{q}}$ with the Geodesic Deviation Equation \citep{vogelsberger_white_2011} and determine the point in time when a particle goes through its first caustic where $\det |d\vec{x}/d\vec{q}| = 0$. Further we use the distortion tensor to get accurate estimates of the stream densities of individual particles $\rho_s = 1/|det \frac{d\vec{x}}{d\vec{q}}|$.

When creating the initial conditions for the simulation we also evaluate the excursion set predictions for the corresponding density field by evaluating the deformation tensor of every particle when using sharp k-space filters with different length scales.

To test whether the Zeldovich approximation and the Triaxial Collapse model give reasonable predictions for the densities of single-stream particles we show in Figure \ref{fig:density_correlation} the measured versus the predicted stream densities of a bootstrapped sample of 1000 particles which have not gone through a caustic in the simulation, and are therefore likely to belong to a single-stream region. Apparently the Zeldovich approximation underpredicts the stream densities of single-stream particles systematically. The triaxial collapse model with the tidal field from linear theory seems to over-predict stream densities, and has a much larger scatter than the pure Zeldovich approximation. This is likely due to the tidal field from linear theory being unreasonably large in the non-linear regime. However, the model with fading tidal field performs significantly better - showing a tight correlation with the right slope, but a slight over-prediction of the densities. There is in principle room for improvement by considering more sophisticated descriptions of the tidal field, but we expect this to be good enough for a first estimate of the density distribution.

\begin{figure}
  \includegraphics[width=\columnwidth]{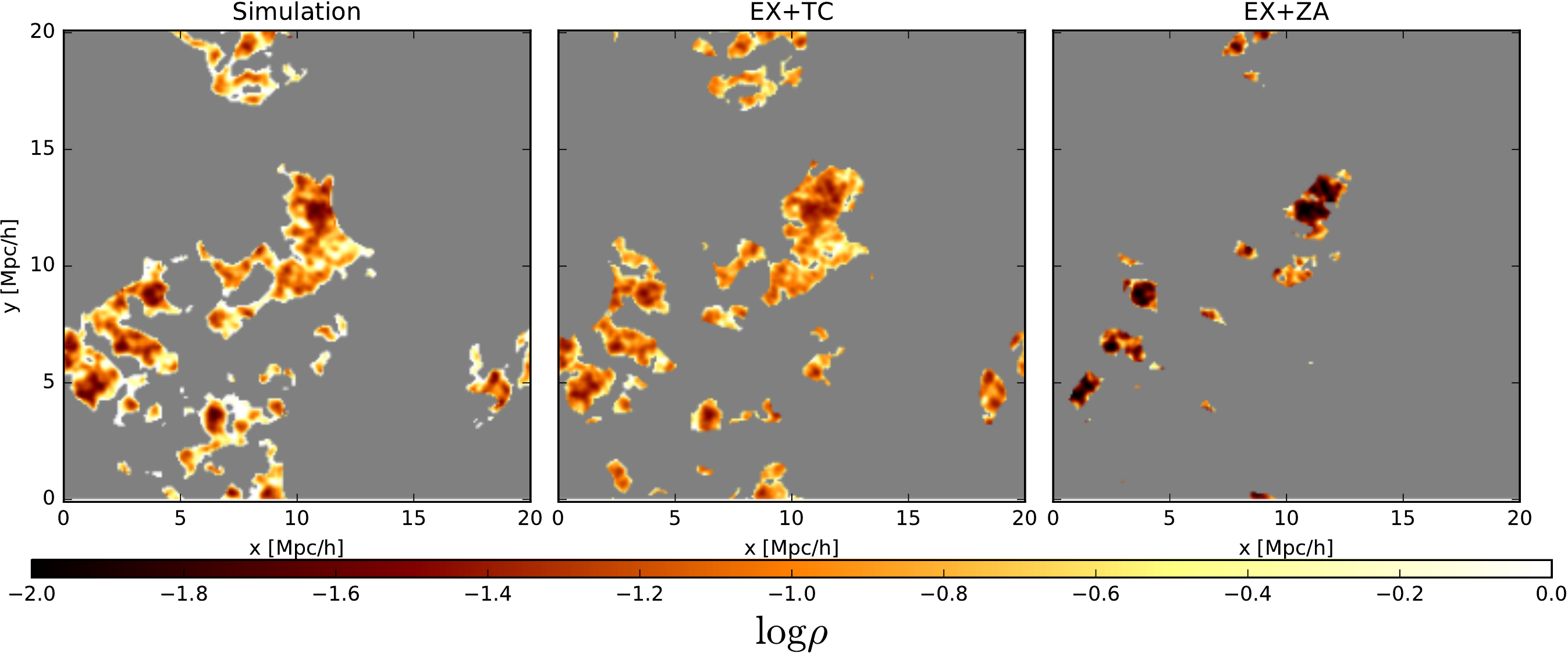}
  \caption{A slice through Lagrangian space showing the stream densities of particles that are considered part of a single-stream region by different approaches. Left: Stream density measured in the simulation for particles that have not yet crossed a caustic. Center: Prediction from the EX+TC approach. Right: Prediction from the EX+ZA approach. The EX+TC agrees fairly well with the simulation. The EX+ZA approach underpredicts the amount of single-stream material and underpredicts the stream densities.}
  \label{fig:lagrangian_class}
\end{figure}
As the second ingredient to the excursion set formalism, the performance of the collapse criterion has to be tested in practice. Therefore we show the in Figure \ref{fig:lagrangian_class} the classification of particles into single-stream (colored) and multi-stream (grey) in a slice through Lagrangian space. Further we color code the predicted stream densities of single-stream particles. The EX+ZA approach apparently generally underpredicts the stream densities of single-stream particles and assigns too little material to single-stream regions. This means that the collapse barrier of the Zeldovich approximation $\lambda_{1,c} = 1$ is too tight in most cases. However, the agreement between simulation and EX+TC is fairly good. This shows that the collapse barrier for single-stream particles should indeed depend on the the two other eigenvalues as in Figure \ref{fig:collapse_barrier} left. If e.g. two axes of a volume element are expanding rapidly ($\lambda_2$, $\lambda_3 \ll 0$), the tidal field needs to act much more strongly than in the Zeldovich case, to bring the smallest axis to collapse.

\begin{figure}
  \includegraphics[width=\columnwidth]{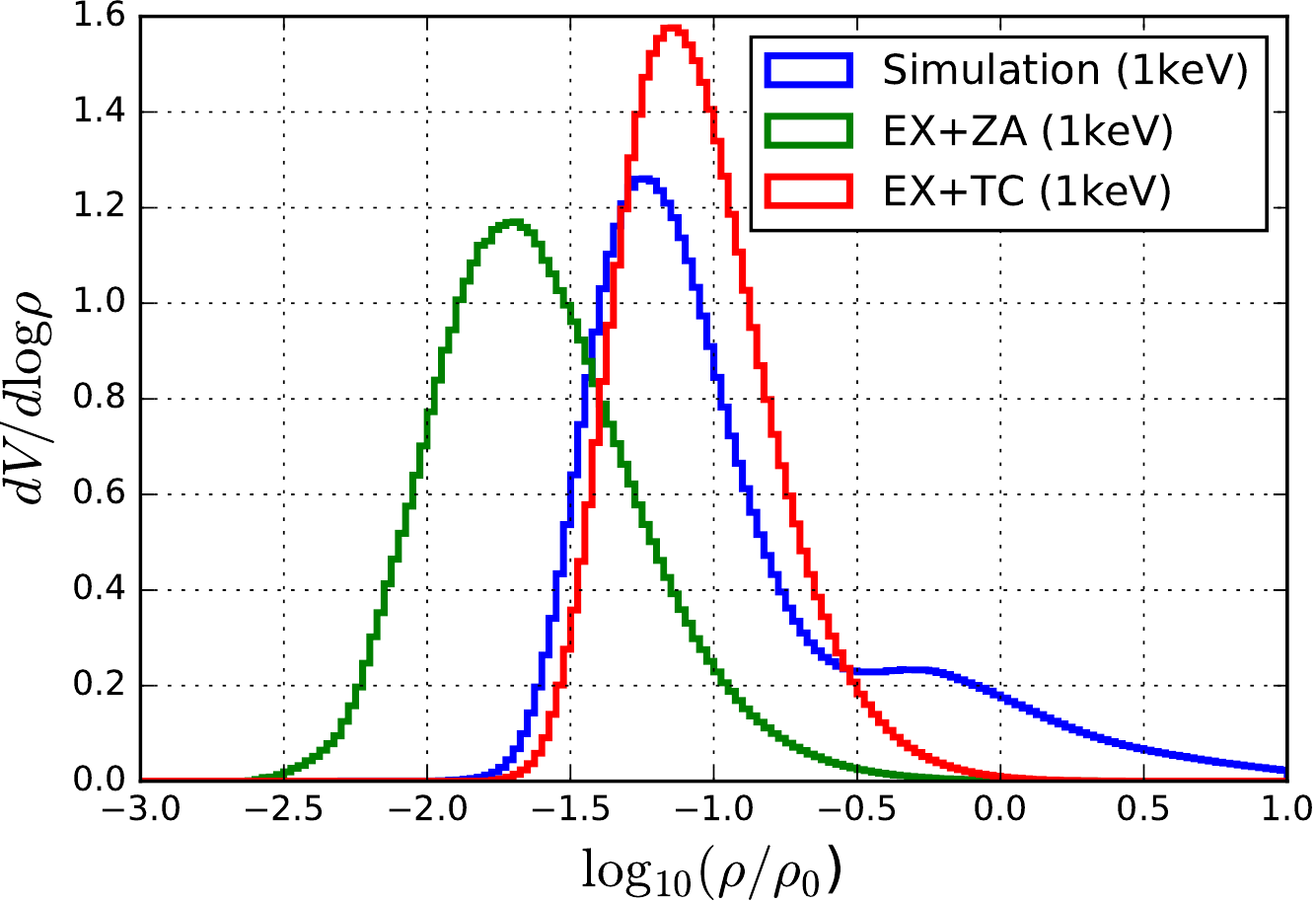}
  \caption{The normalized (volume-weighted) density distribution predictions for single-stream regions by the EX+ZA and EX+TC formalisms in comparison to the measured distribution in a simulation. The high density tail ($\rho \gtrsim 10^{-0.5} \rho_0$) in the density distribution of the simulation mostly originates from multi-stream regions (pancakes, filaments, haloes). These are by construction not represented in the excursion set formalisms.}
  \label{fig:dens_vs_sim}
\end{figure}

As a final benchmark of the scheme, we compare in Figure \ref{fig:dens_vs_sim} the volume weighted density distribution that is predicted by the excursion set formalisms with the distribution that is actually measured in the simulation. The simulation densities are here measured on a mesh in Eulerian space. The high density tail in the simulation is due to multi-stream regions, and it is not expected to be captured by the models. Apparently the EX+ZA formalism strongly underpredicts the densities whereas the EX+TC formalism tends to slightly over-predict densities. Our predictions are not of high accuracy, but they are good enough to provide a reasonable picture and a first estimate of the density distribution. In principle the quality of the triaxial collapse predictions could be improved by finding a more precise description of the tidal field in the non-linear regime. For simplicity however, we stick to our EX+TC model here.
\section{Predictions}
In this section we will use the excursion set formalisms to predict the mass- and volume-fractions of single-stream regions and their density distribution. This distribution should approximately correspond to the volume-weighted density distribution of the universe, except for the high density tail which is dominated by multi-stream regions.

We create a sample of $4 \cdot 10^7$ random walks which we evaluate at
\begin{align}
  \sigma_k &= \frac{k}{n} \sigma_\text{max}
\end{align}
for $n=3000$ steps and $\sigma_\text{max} = 30$. We determine at each step the eigenvalues of the deformation tensor $\lambda_{1-3}$ and test whether one of the Zeldovich or the triaxial collapse criteria is fulfilled. Further we compute several statistics among the particles which are still classified as belonging to a single-stream region. We present these statistics here. $\sigma$ is the only free parameter. Thus, a prediction for a dark matter model today can be read off at its value of $\sigma_\text{Model}$ and a prediction for the model at an earlier time can be simply obtained by reading off at 
\begin{align}
  \sigma (a) = \frac{D(a)}{D(a=1)} \sigma_\text{Model}
\end{align}
For $\sigma$-dependent plots we add the corresponding scale factor of the universe (assuming $\Omega_\Lambda=0.7$ and $\Omega_m=0.3$) for the WIMP model with $\sigma = 23.8$ as an alternative x-scale. We also list some of the predictions for three selected dark matter models in Table \ref{tab:model_data}.
\begin{table*}
  \centering
  \caption{Predictions which are obtained from the excursion set + Zeldovich approximation (EX+ZA) approach and the excursion set + triaxial collapse (EX+TC) approach evaluated for different dark matter models.}
  \label{tab:model_data}
  \begin{tabular}{lcccc}
    \hline
     Dark Matter Model        & WDM (1keV) & WDM (10keV) & WIMP (100GeV) & Axion (10$\mu$eV)  \\
   \hline
    Linear Theory rms density perturbation $\sigma_\text{Model}$ & 4.1        & 7.7        & 23.8  & 32.2       \\
    Single-Stream Mass Fraction (EX+ZA)& $3.98 \cdot 10^{-2}$ & $6.39 \cdot 10^{-3}$ & $2.31 \cdot 10^{-4}$   & $9.54 \cdot 10^{-5}$  \\
    Single-Stream Mass Fraction (EX+TC)& $1.11 \cdot 10^{-1}$ & $3.27 \cdot 10^{-2}$ & $3.09 \cdot 10^{-3}$  & $1.58 \cdot 10^{-3}$   \\
    Median Density (EX+ZA)& $2.18 \cdot 10^{-2}$ & $3.47 \cdot 10^{-3}$ & $1.22 \cdot 10^{-4}$   &  $4.91 \cdot 10^{-5}$  \\
    Median Density (EX+TC)& $8.15 \cdot 10^{-2}$ & $2.93 \cdot 10^{-2}$ & $4.41 \cdot 10^{-3}$  & $2.55 \cdot 10^{-3}$   \\
    Single-Stream Volume (EX+ZA)  & 1.11  &  1.15  &  1.21 & 1.22 \\
    Single-Stream Volume (EX+TC)  & 1.01  &  0.84  &  0.52 &  0.45 \\
   \hline
  \end{tabular}
\end{table*}
\subsection{Mass- and Volume- Fractions}
\begin{figure}
	\includegraphics[width=\columnwidth]{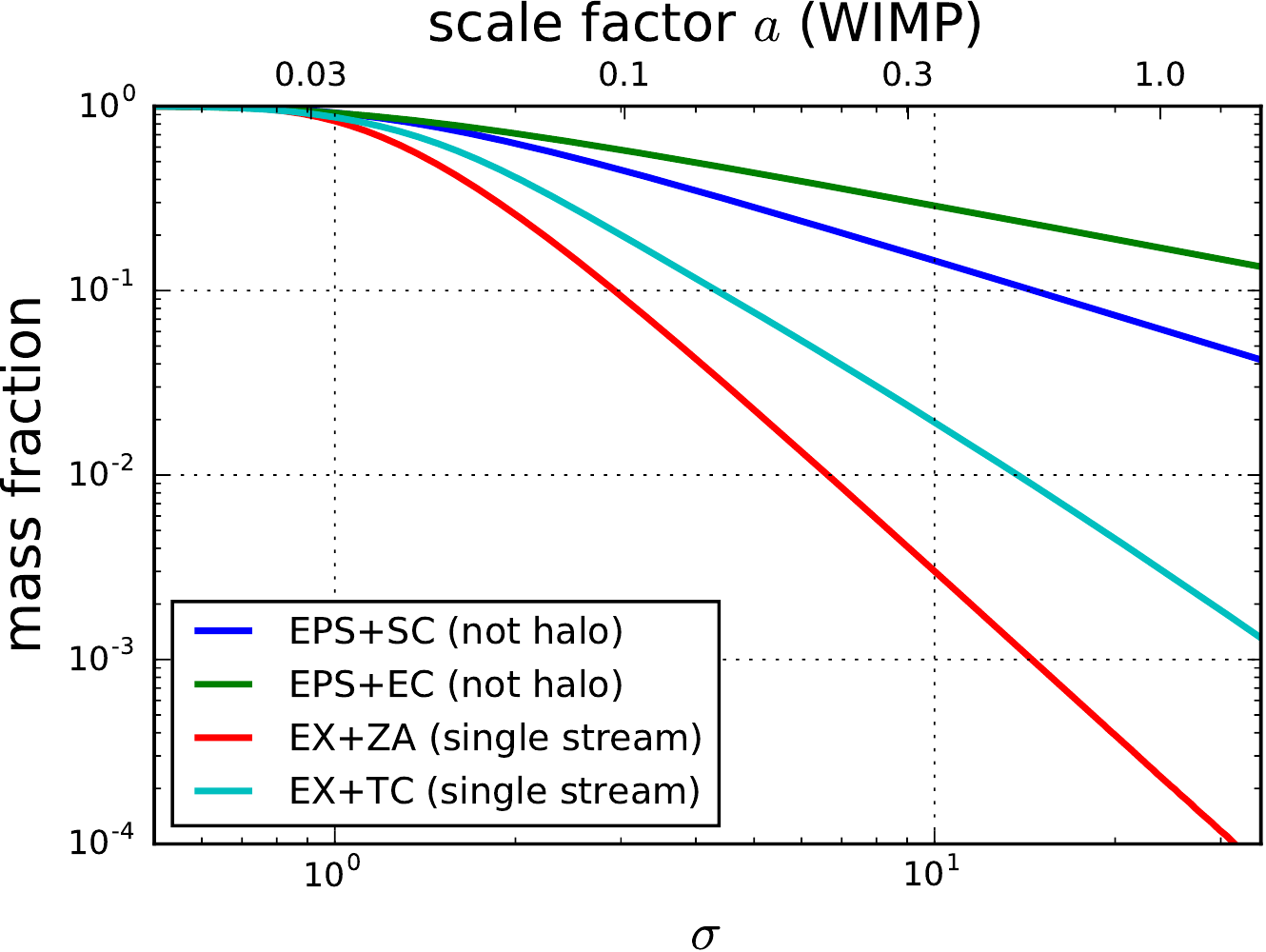}
    \caption{Evolution of the mass fraction of single-stream regions with respect to the total mass (red and cyan). Also indicated the amount of mass that is not part of the halo for the EPS with spherical collapse barrier (blue), and with the moving barrier described in \eqref{eqn:eps_st} (green). At $\sigma=0$ nothing is collapsed and everything is a single-stream region. With increasing density perturbations more and more mass collapses and gets removed from the singlestream regions.}
    \label{fig:void_massfrac}
\end{figure}
In Figure \ref{fig:void_massfrac} we present the mass fraction of single-stream regions with respect to the total mass for the EX+ZA and the EX+TC models. Further we show the amount of material which is predicted not to be part of a halo for the EPS model with barrier $\delta_{sc} = 1.686$ and the EPS model with the moving barrier described in \cite{sheth_tormen_2001} with the parameterization that is used in \cite{angulo_birth_2010}
\begin{align}
  \delta_{ec} (\sigma) = \sqrt{q} \delta_{sc} \left( 1 + \beta \left( \frac{\sigma^2}{q \delta^2_{sc}} \right)^\gamma \right) \label{eqn:eps_st}
\end{align}
with $q = 0.5$, $\beta=0.55$, $\gamma = 0.5$.

 Apparently only a small fraction of the material which is not in haloes ($\sim 10\%$ for a WIMP) is actually in single-stream regions ($0.3\%$ for a WIMP). Most of it is part of filaments and pancakes. Although only a small fraction of the total mass is part of single-stream regions, most of the volume is occupied by them. 

\begin{figure}
	\includegraphics[width=\columnwidth]{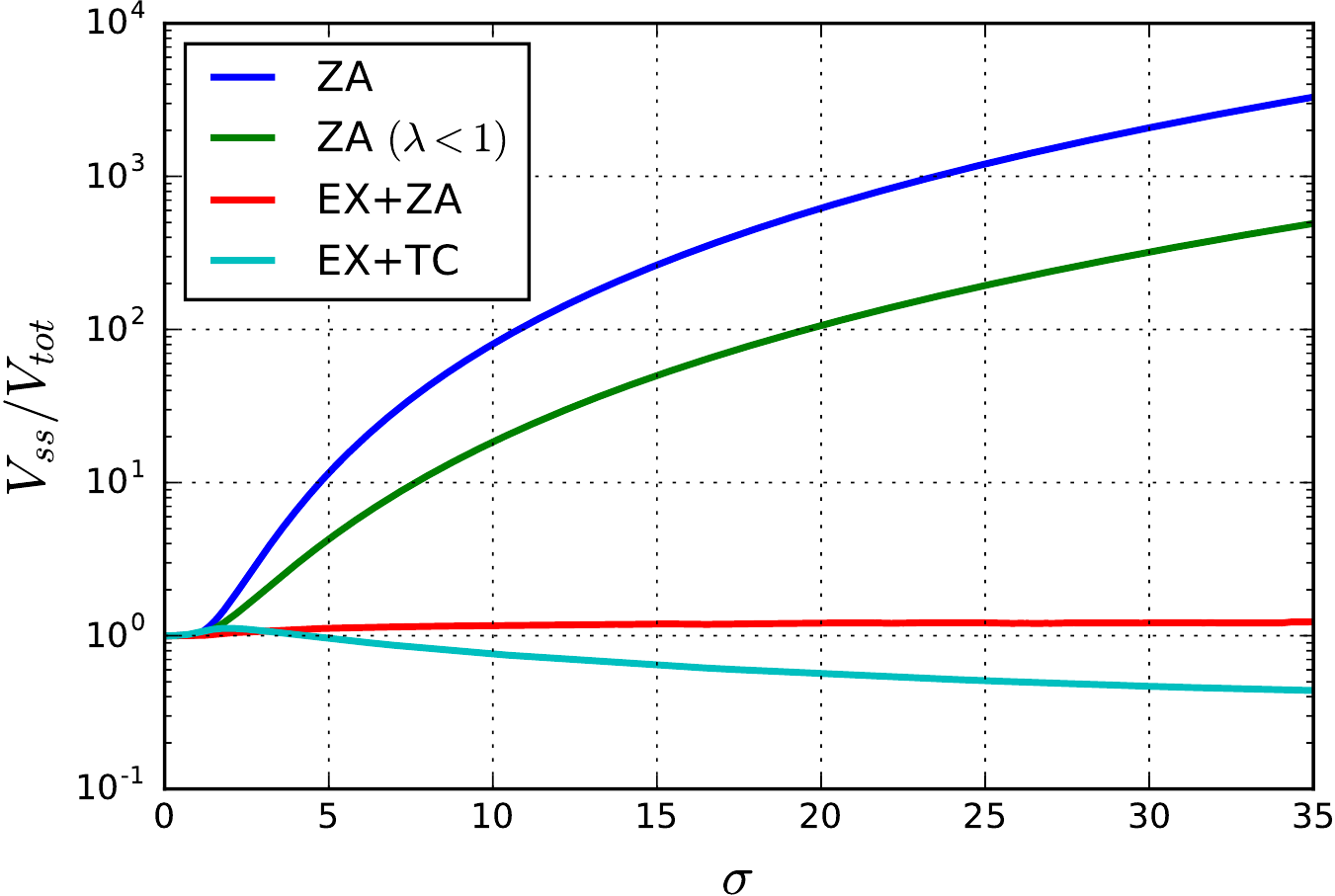}
    \caption{Evolution of the volume fraction of single-stream regions with respect to the total volume. For a pure Zeldovich approximation the volume over counting increases dramatically for $\sigma \gg 1$ and even reaches a factor of $400$ around $\sigma = 30$. With the excursion set formalisms the volume fractions stay of order unity. However, we do not expect the volume fractions to be accurate enough for a quantitative prediction of the single-stream volume. They can merely serve as a consistency check here.}
    \label{fig:void_volumefrac}
\end{figure}
We present the volume fractions that are obtained from the excursion set formalisms in Figure \ref{fig:void_volumefrac}. They are calculated as
\begin{align}
  \frac{V_{ss}}{V_{tot}} = \frac{1}{N_{tot}} \sum_{i \in \text{singlestream}} \frac{\rho_0}{\rho_i} \label{eqn:volumefrac}
\end{align}
where $N_{tot}$ is the total number of particles and the sum only goes over those particles which are considered to be part of single-stream regions. We do not expect these values to be precise, since the density estimates are still relatively crude and the classification operates in Lagrangian Space and some of the particles which are classified as single-stream will actually be in multi-stream regions in Eulerian space. It provides however a benchmark for the scheme, since the total volume of the single-stream regions should be of order unity. This is indeed the case for the EX+ZA scheme which is at about $V_{ss} = 1.19 V_{tot}$ at $\sigma = 23.8$. This is remarkable since the densities tend to be underestimated within this scheme (compare Figure \ref{fig:dens_vs_sim}) and too many particles are classified as collapsed (compare Figure \ref{fig:lagrangian_class}). However these effects seem to balance out so that an Eulerian volume of order unity is achieved. To emphasize this we also show the total volume of the sheet which is obtained in a pure Zeldovich approximation when running the sum of \eqref{eqn:volumefrac} over all particles and when only selecting particles that at the last step of the randomwalk have all eigenvalues with $\lambda < 1$. In both of these cases the volume is overestimated by several orders of magnitude. This is due to shells expanding after shell-crossing without any de-accelerating forces as we have already seen on the left side of Figure \ref{fig:zeldovich_slice}. The EX+TC scheme seems to underpredict the volume a bit which leads to $V_{ss} \sim 0.52 V_{tot}$ at $\sigma=23.8$. This is probably due to the densities being slightly overestimated as we have seen in Figure \ref{fig:dens_vs_sim}. However, the deviation is still well below factor of two and therefore unproblematic for our current approximate estimates.

\subsection{Density Distribution}
\begin{figure}
	\includegraphics[width=\columnwidth]{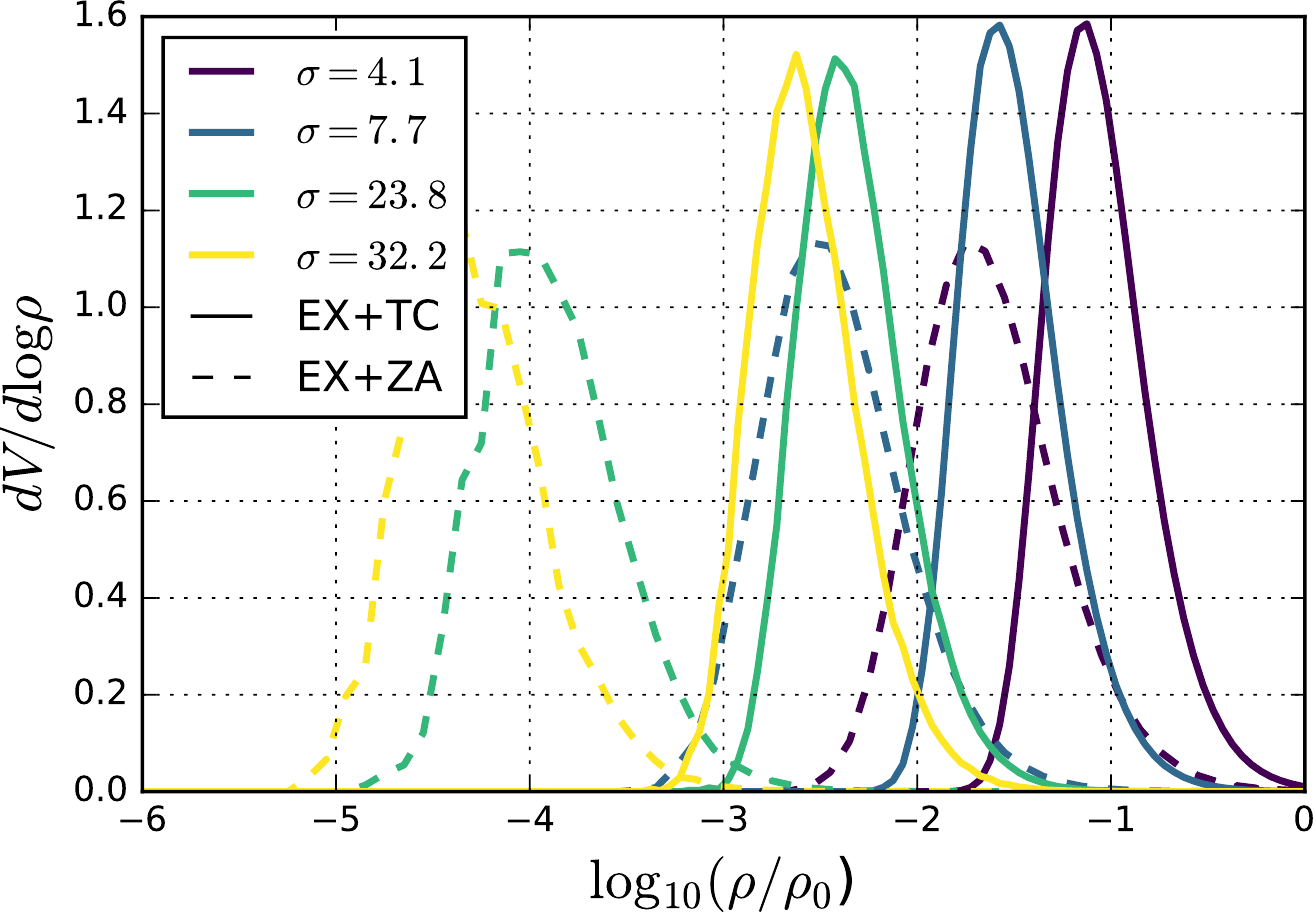}
    \caption{Volume weighted density histogram for the single-stream regions for four different dark matter models: WDM (1keV) in black, WDM (10keV) in blue, WIMP (100GeV) in green and an axion ($10 \mu$eV) model in yellow. All distributions are normalized to 1. The dashed lines represent the EX+ZA prediction and the solid lines the EX+TC model prediction. We expect the true distributions to lie a bit to the left of the EX+TC predictions.}
    \label{fig:void_denshist_volume}
\end{figure}
We present the volume weighted density distribution for the single-stream regions of the four considered dark matter models in Figure \ref{fig:void_denshist_volume}. For each model we present the EX+ZA and the EX+TC prediction. From Figure \ref{fig:dens_vs_sim} we would expect that the EX+ZA scheme significantly underpredicts the densities and the EX+TC distribution slightly over predicts them. The true density distribution probably lies slightly to the left of the EX+TC distributions. Interestingly the shape of the distributions changes little as the variance of the density field increases; rather their center just shifts to lower densities. We list the median densities of the distributions in Table \ref{tab:model_data}. The EX+TC scheme predicts for a WIMP of mass 100 GeV a median density of $\sim 4 \cdot 10^{-3} \rho_0$. Following our argumentation from above, if the dark matter is indeed such a WIMP, this should roughly be the \emph{median density of the universe}.

\begin{figure}
    \includegraphics[width=\columnwidth]{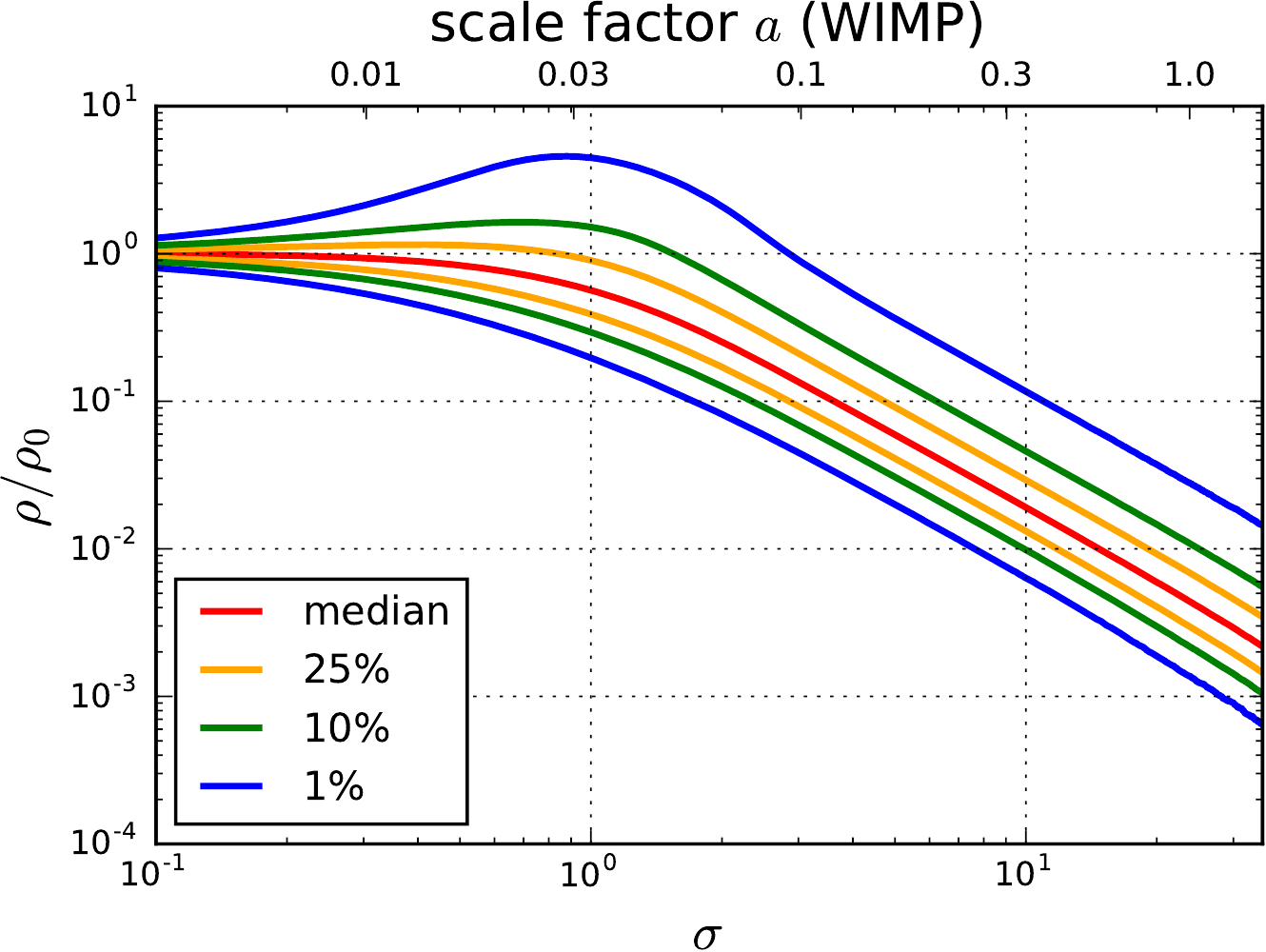}
    \caption{Quantiles of the volume weighted density distribution for the EX+TC model as a function of the linear theory rms density perturbation $\sigma$. The upper x-scale shows the corresponding scale factors for a $\Lambda$CDM cosmology with a 100 GeV WIMP as dark matter. While for $\sigma \lesssim 0.2$ the distribution is symmetric around the mean density $\rho = \rho_0$, for $\sigma \gg 1$ the density distribution is dominated by the under dense regions.}
    \label{fig:dens_quantiles_volume}
\end{figure}

Further we show the evolution of the quantiles of the distribution for the EX+TC scheme in Figure \ref{fig:dens_quantiles_volume}. Note how the the normal distribution around $\rho = \rho_0$ is reproduced in the linear regime for $\sigma \lesssim 0.2$. In this regime the median density is approximately the mean density $\rho_0$, since over- and underdense regions enter the distribution roughly with the same weight. For the WIMP model the volume density distribution starts deviating significantly from the linear case for $a \gtrsim 0.01$. In the strongly non-linear regime $\sigma \gg 1$ the median density roughly scales like $\rho \propto \sigma^{-3/2}$. At that point the volume density distribution is completely asymmetric, and dominated by underdense regions.

\section{Percolation} \label{sec:percolation}

Whereas above we focused on predicting the density distribution, and the volume- and mass-fractions of single-stream regions, we will try to find out more about their topology in this section. A profound question is whether it actually makes sense to speak of individual single-stream regions, or whether almost all single-stream volume is connected in one universe filling complex. That is, \emph{do single-stream regions percolate?}

\subsection{Previous Work}

 This question has already been investigated by \citet{falck_neyrinck_2015} who found that the single-stream regions in their simulations do indeed percolate. 
 
They use simulations of resolution up to $L/N^{1/3} = 0.2 h^{-1} \text{Mpc}$ (boxsize over particles per dimension). As can be read off from Figure \ref{fig:darkmatter_models} this corresponds to a resolved rms density perturbation of approximately $\sigma(k_\text{Ny} \sim 16  h \text{Mpc}^{-1}) \sim 5$. They use the \verb'ORIGAMI' \citep{falck_2012} method to identify single-stream regions in their simulations. Each particle is tested for shell-crossing along three orthogonal Lagrangian axes. A single-stream particle is then a particle with no shell-crossing along any axis. 

\citet{falck_neyrinck_2015} find that in all their tests the single-stream regions appear to percolate. Further they find that the mass fraction of single-stream regions decreases when going to higher resolution, whereas the volume fraction does not decrease significantly. From this, they infer that single-stream regions may percolate even in the case of infinite resolution.

We argue here, however that the fact that the mass fraction of single-stream regions decreases significantly with increasing resolution (or equivalently: with increasing $\sigma$), makes it quite unlikely that single-stream regions percolate: In single-stream regions the mapping from Lagrangian space $\vec{q}$ to Eulerian space $\vec{x}$ is one to one (per definition). Therefore if single-stream regions form an infinitely large percolating structure in Eulerian space, they also form an infinitely large percolating structure in Lagrangian space. Now, at infinite resolution only a tiny fraction of the mass is part of single-stream regions (e.g. $0.3 \%$ for a WIMP). That means single-stream regions would have to be percolating in Lagrangian space at a (Lagrangian) volume fraction of $\sim 0.3\%$ (for a WIMP).

While we agree that single-stream regions are likely to percolate as long as they contain significant fractions of the total mass, we expect them to stop percolating when they reach a low enough mass fraction. As already shown in Figure \ref{fig:void_massfrac} the mass fraction is only dependent on $\sigma$, and therefore we ask the question: \emph{Do single-stream regions always percolate, and if not, at which value of $\sigma$ do they stop percolating?}

\subsection{Percolation with the Excursion Set formalism}

\begin{figure}
  \includegraphics[width=\columnwidth]{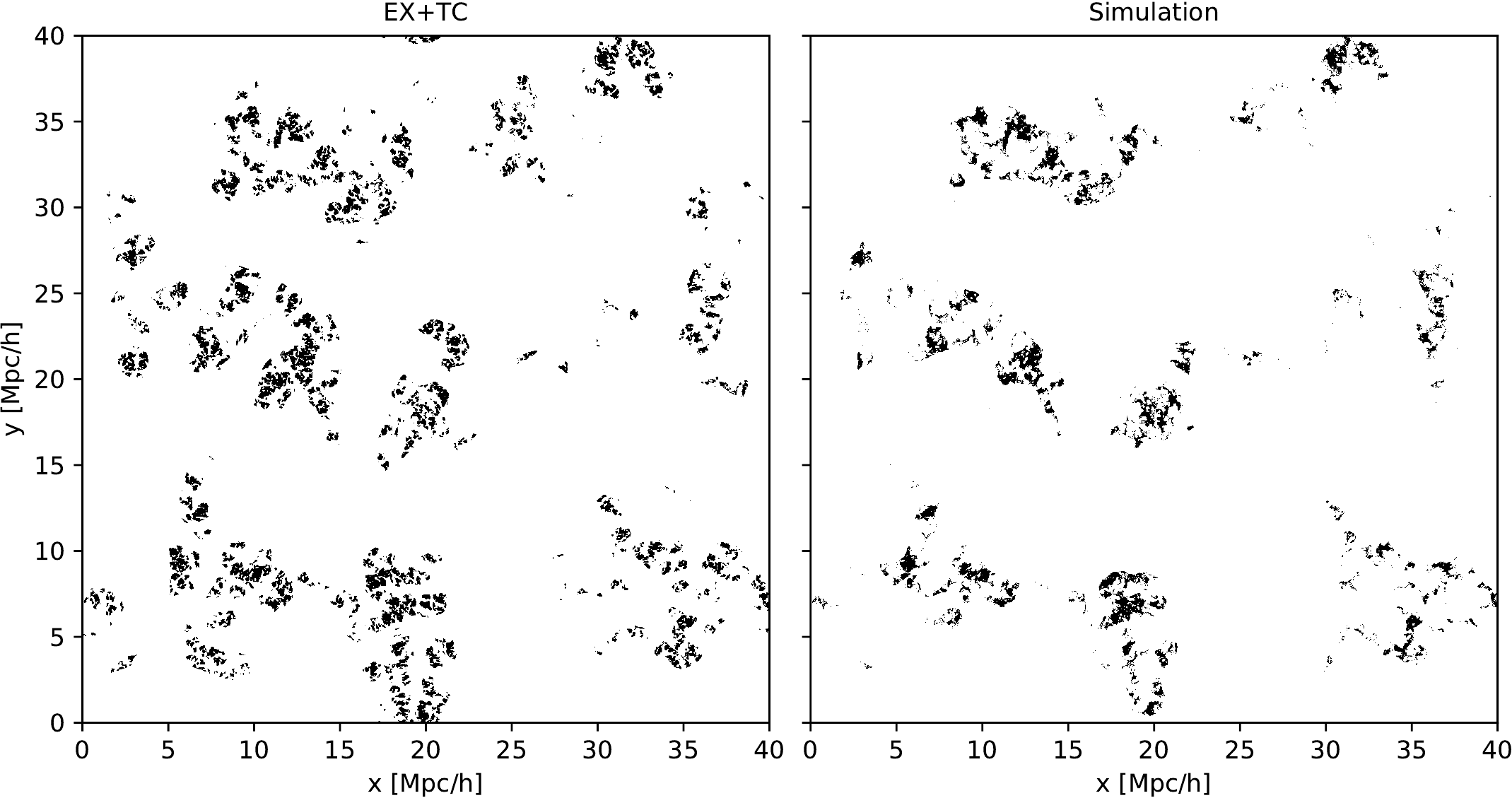}
  \caption{Lagrangian slice classifying particles into single-stream (black) and multistream (white) for a comsological box with $\sigma=6.4$. Left: Classification according to the EX+TC excursion set prediction. Right: Classification inferred from a simulation. The particles are classified according to the nearest grid point in Eulerian Space as explained in section \ref{sec:perclag}. In both cases the single-stream regions do not percolate.}
  \label{fig:lagrangian_ss}
\end{figure}

We try to search the percolation threshold of single-stream regions with the EX+TC scheme here. As already shown in Figure \ref{fig:lagrangian_class}, the EX+TC scheme performs fairly well in predicting which particles are going to become part of a single-stream region. Therefore we can use it to test for the percolation of single-stream regions in Lagrangian space to get an estimate of the percolation threshold. It is computationally far cheaper to evaluate than to run a whole simulation.

We set up a Lagrangian mesh of $1024^3$ particles within a periodic box of a width of $L = 40 \text{Mpc/h}$. We use the same power spectrum as shown in Figure \ref{fig:darkmatter_models} with a sharp k-space cutoff at  $k_c = \frac{2 \pi}{0.1 Mpc} = 63 \text{Mpc}^{-1}$  leading to $\sigma_0 = 6.4$. For each particle the EX+TC scheme is used to determine the growth factor at collapse $D_\text{col}$. In Figure \ref{fig:lagrangian_ss} (left) we show a thin slice through Lagrangian space for particles which have $D_\text{col} > D(a=1) = 1$, i.e. particles which are single-stream at $a=1$.

We test for different growth factors $D_\text{thresh}$ whether the boolean grid defined by $D_\text{col} > D_\text{thresh}$ percolates. To test for percolation, we determine all connected single-stream regions in Lagrangian space, linking the cells over the faces to their 6 nearest Langrangian neighbours. A region percolates if there is a path between any cell and any of its periodic replications that does not exit the connected component.

 We find the single-stream regions to percolate only for $D_\text{thresh} < 0.77$ - corresponding to $D \cdot \sigma_0 = 4.9$ and a single-stream mass fraction of $8.0\%$. Although there might be a dependence of the percolation threshold on the precise shape of the power spectrum and also a residual realisation dependence (due to the small box size), we expect these influences to be small in comparison to the importance of the mass fraction which exclusively depends on the value of $D \cdot \sigma_0$. 

So, assuming that these factors are small, we predict with the EX+TC scheme that single-stream regions do not percolate if 
\begin{align}
  \sigma_\text{DM}(a) \gtrsim 5  \text{  .}
\end{align}
It is an unfortunate coincidence that this value is so close to the resolution limit of  \citet{falck_neyrinck_2015}. Possibly their single-stream regions are not far away from ceasing to percolate - at one or two resolution levels higher they might have found them to no longer percolate.

However, it is not clear whether the EX+TC prediction is accurate enough for this to be a reliable conclusion. Fortunately $\sigma \gtrsim 5$ is still in a regime which is accessible through classical dark matter simulations so that it can be tested numerically.
\subsection{Percolation in a Simulation}

\begin{figure*}
  \includegraphics[width=\textwidth]{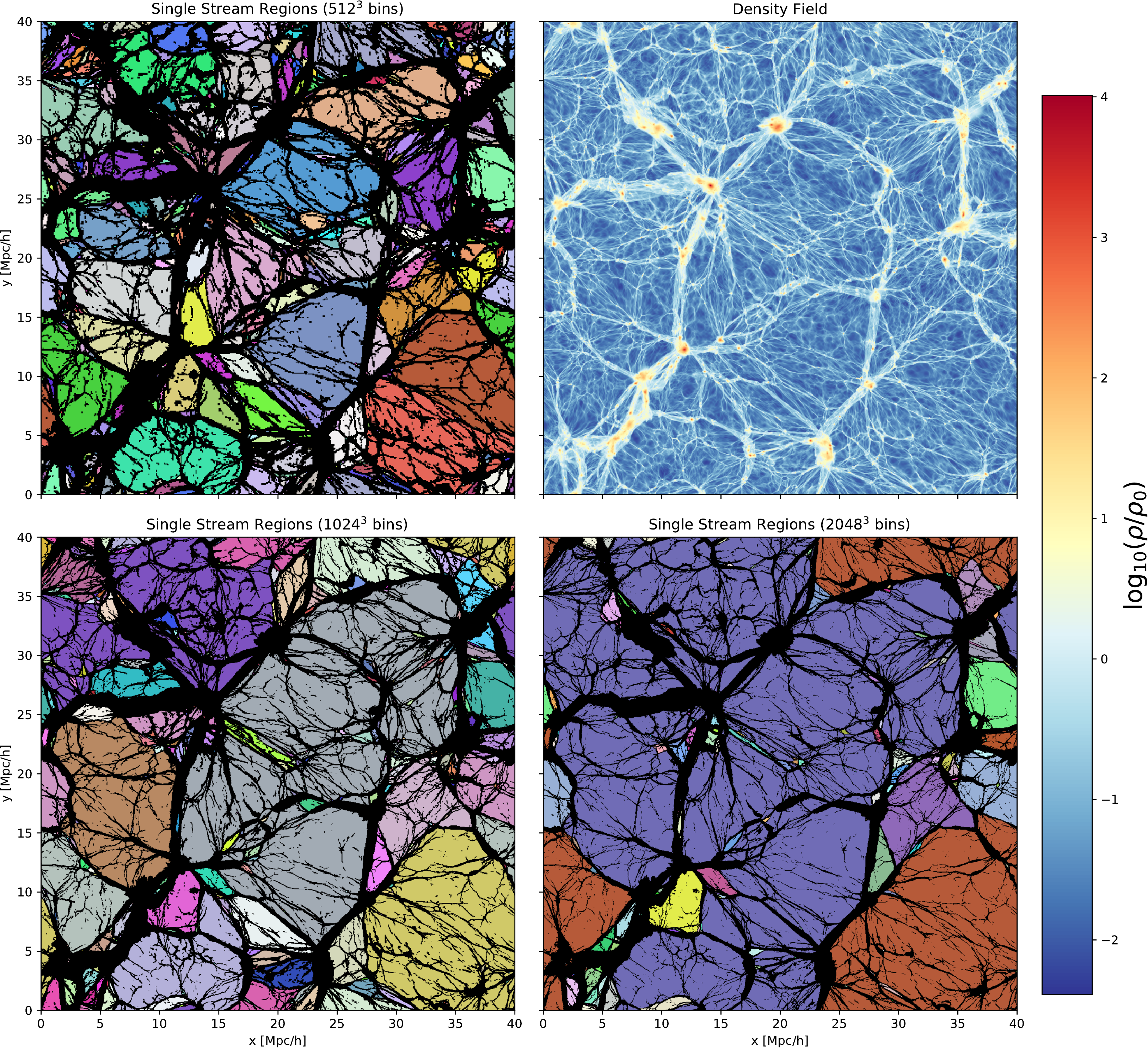}
  \caption{A thin slice (width 40 kpc) through a cosmological $\Lambda$CDM simulation. Top right: density field (logarithmic).  Other panels: Classification into multi-stream regions (in black) and single-stream regions (color) for different different grid resolutions: top left $512^3$ bins, bottom left $1024^3$ and bottom right $2048^3$. Each distinct single-stream region is assigned a random color. Many of the connected regions do not appear to be connected within this slice, but are connected through the third dimension. In the cases with $512^3$ and $1024^3$ bins the single-stream regions do note percolate. In the $2048^3$ case there is one region (dark blue) which percolates in the y-dimension, but not in the x- and z-dimensions. Note that since this is a thin slice through Eulerian space, most structures that appear string-like are slices through pancakes. We provide movies that scroll through the z-coordinate of these slices under \href{http://wwwmpa.mpa-garching.mpg.de/paper/singlestream2017/percolation.html}{wwwmpa.mpa-garching.mpg.de/paper/singlestream2017/percolation.html}.}
  \label{fig:dens_singlestream}
\end{figure*}

We attempt to test the prediction that single-stream regions do not percolate for  $\sigma \gtrsim 5$ with a simulation. Therefore we use the same particle grid that has been used for the EX+TC calculation ($\sigma=6.4$) to create initial conditions with the Zeldovich approximation at a scale factor of $a=0.01$. We run a classical N-Body simulation using the Planck $\Lambda$CDM cosmology $\Omega_m = 0.30$, $\Omega_\Lambda = 0.69$ and $h=0.68$ \citep{planck_2016} up to a scale factor of $a=1$. Then we use only the positions of the particles at the final time to calculate which regions of the space are single-stream regions and which are multi-stream regions. We give a detailed description of the algorithm we use in Appendix \ref{app:trigores}. A short summary follows:

\subsubsection{Single-Stream Classification}
\label{sec:ssclass}

By phase space interpolation techniques \citep{abel_2012, shandarin_2012, hahn_angulo_2016} we create from the $1024^3$ original particles a much larger number ($32768^3$) of re-sampled particles which we bin onto a $512^3$, a $1024^3$ and a $2048^3$ cubic mesh to infer a high quality density field. Additionally, we determine for the re-sampled particles the determinant of the real space distortion tensor
\begin{align}
  D_{xq} &= \det \left(\frac{\partial \vec{x}}{\partial \vec{q}} \right)
\end{align}
where $\vec{x}$ are Eulerian and $\vec{q}$ are Lagrangian coordinates. We determine all bins which contain any re-sampled particles that have a negative determinant and classify them as multi-stream regions. Note that this works, since all multi-stream regions contain streams which have gone through at least one caustic, and further, every time a particle goes through a caustic, the determinant of the distortion tensor flips its sign - starting with a positive sign initially \citep{vogelsberger_white_2008, vogelsberger_white_2011, shandarin_medvedev_2014}. We give a more detailed description of this algorithm in Appendix \ref{app:trigores}. Note that this way of determining single-stream regions is very robust and simple to implement. It can in principle be used on every cosmological dark matter simulation which uses grid like initial conditions.

We show a slice through the determined density and single-stream fields in Figure \ref{fig:dens_singlestream}. We color distinct single-stream regions in random colors. Every bin is linked with its 6 nearest neighbours (no diagonals). Additionally we provide a set of movies and additional material at the following address: \href{http://wwwmpa.mpa-garching.mpg.de/paper/singlestream2017}{wwwmpa.mpa-garching.mpg.de/paper/singlestream2017}. This includes, for example, movies which scroll through the z-coordinate over time. Visualizing the whole three dimensional volume this way helps to understand how regions are connected.
 
\subsubsection{Percolation in Eulerian Space}

\begin{figure}
  \includegraphics[width=\columnwidth]{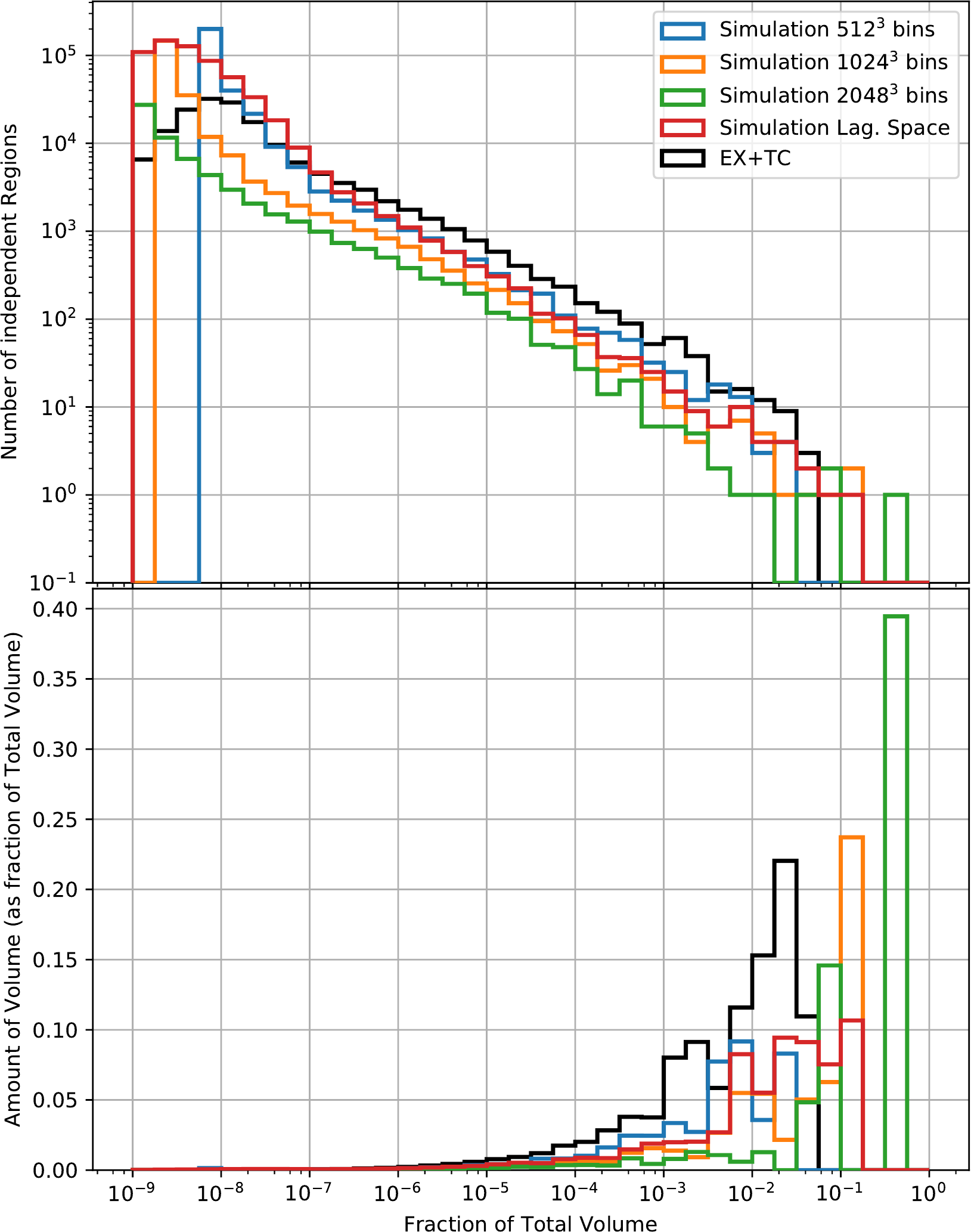}
  \caption{Top: Histogram of the Number of independent single-stream regions of a given size. The x-scale is given as fraction of the total volume $V_\text{tot} = (40 \text{Mpc}/h)^3$. Bottom: Volume weighted histogram. Apparently in the case with 2048$^3$ bins there is a single-stream region which takes about $40\%$ of the total volume and which percolates in one dimension. The areas under the curves correspond to the total single-stream volume and are listed in Table \ref{tab:percolation-data}.}
  \label{fig:ss-histogram}
\end{figure}

\begin{table*}
  \centering
  \caption{Volume and mass-fractions of single-stream regions. Note that for the excursion set case (EX+TC) the numbers given here are for the Eulerian volumes corresponding to connected zero caustic crossing regions in Lagrangian space. These are expected to be larger than Eulerian single-stream regions and to sum to somewhat more than the total (Eulerian) volume.}
  \label{tab:percolation-data}
  \begin{tabular}{l c c c c c c}
    \hline
    bins & binwidth & total ss. vol. & total ss. mass & largest component vol. & second largest component vol. & percolation \\
    \hline
     512$^3$   & 78 kpc/$h$ & 46.5\%  & 1.5 \% & 2.1 \% & 2.3 \% & none\\
     1024$^3$ & 39 kpc/$h$ & 59.6\%  & 2.2 \% & 12.4 \% & 11.3 \% & none\\
     2048$^3$ & 20 kpc/$h$ & 67.4\%  & 2.7 \% & 39.5 \% & 7.5 \% & 1D \\
     Lag. Space  &             & 66.2\%  & 2.8 \% & 10.7 \% & 7.5 \% & none\\
     EX+TC        &             & 102\%   & 5.1 \% & 8.5 \%   & 8.4 \%  & none\\
   \hline
  \end{tabular}
\end{table*}

In the case with $512^3$ bins we clearly find many individual single-stream regions that do not percolate. However, if we increase the number of bins and therefore decrease the bin size, many of the originally independent regions connect together to larger regions. This leads to much larger typical single-stream regions in the case with $1024^3$ bins and even percolation in one dimension (as we will discuss below) in the case with $2048^3$ bins. That means that single-stream regions with diameters of a few Mpc get linked together through gaps smaller than a binwidth of $39$kpc.

This can also be seen in the volume histograms of single-stream regions which we show in Figure \ref{fig:ss-histogram}. In all three cases, as well as in the excursion set prediction, the number count of single-stream regions follows a power law with a slope $n(V) \propto V^{-0.5}$ up to a largest region. However, while there seems to be no preferred largest region in the $512^3$ and the $1024^3$ bins cases, the largest region in the $2048^3$ bins case makes up $39.5\%$ of the total volume whereas the second largest makes up only $7.5\%$. We list these values among other relevant mass- and volume-fractions in Table \ref{tab:percolation-data}. 

It is hard to conclude from this alone whether single-stream regions percolate or not. The largest component clearly stands out in the $2048^3$ bin case. However, if it was percolating, it would be surprising that it still only makes up a bit more than half of the total single-stream volume of $67.4\%$.

Through a more sophisticated percolation test we find that the largest region in the $2048^3$ bin case is percolating in the y-dimension, but not in the x- and z-dimension. That means the region together with its periodic replications would form an infinitely large string-like structure along the y-dimension, but would not be connected to its periodic images in the $x-$ and $z-$ directions. Percolation in a single dimension is practically impossible in an infinite universe, and the fact that we find it in this periodic box, shows that we are limited by finite size effects.

\subsubsection{Resolution Effects}

It is somewhat surprising that the connectivity of the single-stream regions depends so strongly on the resolution of the mesh. The reason for this could be either (1) that there are tiny holes in the multi-stream regions that only get resolved at higher mesh resolutions or (2) that the likelihood of linking regions together through numerical artefacts increases strongly with the mesh resolution.

We find some evidence that (2) might be the case here. While we use exclusively a resampling with $32768^3$ particles in the plots in this section, we tested also how the resampling resolution affects the results. Generally we find that the single-stream field is very well converged with the used resampling resolution. For example when switching from $16384^3$ to $32768^3$ resampled particles, the volume fraction of single-stream regions in the $1024^3$ bins case only decreases by $0.3\%$ from $59.9\%$ to $59.6\%$. While this proves that there is only a small uncertainty in the classification of the cells, such a small difference can lead to major differences in the linked regions. For example the same resolution switch changed the volume of the largest connected region from $22.7\%$ to $12.4\%$ - apparently by disconnecting the largest connected component into two independent subcomponents. Similar disconnections might be expected for the $2048^3$ bins case when further increasing the resampling resolution.

We speculate that the dependence of the connectivity on such tiny details is probably due to the large surface area of the single-stream - multi-stream intersection and the small thickness of the multi-stream regions. The large surface area leads to a large number of boundary pixels that need to be classified correctly, while the small thickness of multi-stream regions leads to a larger chance of single-stream regions connecting through misclassified pixels. When increasing the number of bins at constant resampling resolution, the number of surface area pixels increases dramatically, and the chance of misclassifying individual pixels increases slightly - thereby  increasing the chance of non-physical connections significantly.

\subsubsection{Percolation in Lagrangian Space} \label{sec:perclag}

These resolution issues can in principle be solved by using even higher resampling resolutions so that the chances of misclassifications diminish. However, a cheaper alternative is to map the single-stream field into Lagrangian space, and link resolution elements there. In the continuum limit, linking the single-stream regions in Lagrangian space or in Eulerian space should be equivalent, since the mapping from Lagrangian to Eulerian space is one-to-one and continuous in single-stream regions. However, at finite resolution the chances of linking together unconnected regions due to misclassified cells is much smaller in Lagrangian space. This is because the disconnecting multi-stream regions are much larger in Lagrangian space, and the number of pixels that define the intersections is much smaller. This becomes obvious when comparing Figure  \ref{fig:lagrangian_ss} with Figure \ref{fig:dens_singlestream}.

We classify each particle as single-stream or multi-stream according to the class of the nearest grid point from the $2048^3$ bin Eulerian single-stream field. We show a slice through the classification in Lagrangian space in the right panel of Figure \ref{fig:lagrangian_ss}. Then we determine the connected components\footnote{We provide movies showing these in 3D at \href{http://wwwmpa.mpa-garching.mpg.de/paper/singlestream2017/percolation.html}{wwwmpa.mpa-garching.mpg.de/paper/singlestream2017/percolation.html}. } and determine their volumes by weighting with $1/\rho_s$ where $\rho_s$ is the Eulerian density at the particle positions. We provide the volume histogram of the single-stream regions as the red line in Figure \ref{fig:ss-histogram}.

We find that the single-stream regions do not percolate in Lagrangian space. The volume distribution of single-stream regions appears to be relatively similar to the one in the $1024^3$ bin case. This could possibly mean that resolution effects are best under control in that case. 

However, while we think that the test for percolation in Lagrangian space should be more stable against resolution effects than the Eulerian one, we are still affected by finite size effects here. The largest single-stream region still makes up $10.7 \%$ of the total volume, and the volume histograms look quite noisy due to the low number statistics. In principle a simulation with larger boxsize could help out. However, at the same time it appears to be important that the resolution is sufficiently high to capture the power spectrum far enough ($\sigma \gg 5$) and to resolve small features ($20$-$40$kpc) in the single-stream field. This would make a better test simulation computationally difficult.

\subsection{Do Single-Stream Regions percolate?}

Unlike \citet{falck_neyrinck_2015} we do not find any case where single-stream regions clearly percolate. This may reflect differences in methodology, or may be because we test for percolation at a much higher variance of the density field $\sigma \sim 6.4$ where the mass fraction of single-stream regions is significantly smaller. While we find a case which gets close to full three dimensional percolation ($2048^3$ bins in Eulerian space), we suspect that resolution effects are enhancing connectivity here. A more robust test in Lagrangian space shows no sign of percolation. However, the single-stream regions that we find still occupy a major fraction of the box volume, and thus are still arguably quite close to percolation. A larger simulation could be made to test for finite size effects, but would be computationally expensive. Nevertheless, we suspect that single-stream regions do not percolate in the continuum limit of cold dark matter. The more small scale power is resolved, the less mass remains in the single-stream regions, the smaller the volume of Lagrangian space they occupy,  and the less likely percolation becomes.

\section{Conclusions}
Simulating the unsmoothed dark matter distribution in the deeply non-linear regime is not yet possible for WIMP-like dark matter models, which have a high variance and require resolving many orders of magnitude in spatial scale. We propose here an excursion set formalism to get an estimate for the volume weighted density distribution. This allows us to estimate the median density of the universe $4 \cdot 10^{-3} \rho_0$ and the fraction of mass in single-stream regions $3 \cdot 10^{-3}$ if dark matter is a 100 GeV WIMP. The only parameter in the excursion set model is the rms relative density perturbation $\sigma$ which is directly related to the properties of the specific dark matter model considered. Thus, in principle, a measurement of e.g. the \emph{median density of the universe} would provide information about the nature of the dark matter particle. Unfortunately this is unlikely ever to be possible, since all measurements of dark matter are indirect and involve smoothing on a relatively large scale.

The proposed excursion set formalism gives a reasonable qualitative picture and an first estimate for the quantitative properties of single-stream regions. However, there is still a lot of room for improvement. A significant limitation comes from our heuristic assumptions about the tidal field in the non-linear regime. A more sophisticated examination of the statistics of the tidal field within simulations could lead to a superior model. Further it would be straightforward to go to a full tensor-valued evolution model (given by the Geodesic Deviation Equation). This would require dealing with changes in the orientation of the tidal field.

Further the excursion set formalism predicts that single-stream regions do not percolate if the resolved linear rms density perturbation satisfies $\sigma \gtrsim 5$. We suspect that previous studies \citep{falck_neyrinck_2015} did not reach high enough resolution to reach the state where single-stream regions no longer percolate. We attempted to test this prediction with a smaller, but higher resolution  simulation with $\sigma \sim 6.5$. This simulation shows many small distinct single-stream regions, but their distribution depends strongly on the choice of resolution parameters, even if the classification of individual cells into single and multi-stream regions seems to be converged reasonably well. We suspect that percolation detection in Eulerian space is susceptible to numerical artefacts, since the surface area of the single-stream regions is enormous whereas the thickness of the disconnecting multi-stream regions is relatively small. A percolation test in Lagrangian space, which should be numerically more robust, shows no signs of percolation. 

It would be interesting to run a larger simulation with the same mass resolution to get a better statistical grasp of the volume distribution of single-stream regions, and to test this on a scale, where finite size effects do not matter anymore. Our expectation is that single-stream regions will not percolate in the continuum limit of cold dark matter. At $\sigma \sim 6.4$ we already observe a state where single-stream regions are typically disconnected. Cold dark matter has a much larger variance of the density field (e.g. $\sigma = 23.8$ for a 100GeV WIMP) and therefore a significantly smaller mass fraction in single-stream regions. As a result, these occupy a substantially smaller volume in Lagrangian space and hence seem much less likely to percolate than in the (already non-percolating) simulation we have been able to carry out.

\section*{Acknowledgements}
The authors thank Mark Vogelsberger for the access to the geodesic deviation equation code. JS thanks Oliver Hahn and Raul Angulo for helpful discussions and for providing a motivating and inspiring atmosphere within the collaboration. 



\bibliography{../bibliography} 
\bibliographystyle{mnras}


\appendix

\section{Derivation of the Triaxial Collapse Model} \label{app:triaxial_collapse}
\subsection{The Geodesic Deviation Equation}
In the comoving frame the equations of motion for the coordinates of a test particle can be written as
\begin{align}
  \dot{x_i} &= p_i a^{-2} \\
  \dot{v_i} &= - \partial_i \phi(\vec{x}) a^{-1}
\end{align}
where $v_i / a$ are the peculiar velocities, $x_i a$ are physical coordinates and $\phi$ is the comoving potential which follows the Poisson equation
\begin{align}
  \nabla^2 \phi &= 4\pi G \rho_{bg} \delta \\
  \delta            &= \frac{\rho - \rho_{bg}}{\rho_{bg}}
\end{align}
where $\rho_{bg}$ is the (time dependent) background density and  derivatives are taken with respect to the comoving coordinates.

For an infinitesimal small volume element around a particle the distortion with respect to the homogeneous initial state (Lagrangian Space) is described by the distortion tensor
\begin{align}
   D_{xq} := \frac{d\vec{x}}{d\vec{q}} \\
   D_{vq} := \frac{d\vec{v}}{d\vec{q}}
\end{align}
where $\vec{q}$ are Lagrangian coordinates so that $\vec{x}(a \rightarrow 0) = \vec{q}$. The evolution equations of the distortion tensor can be derived as follows:

\begin{align}
  \frac{d}{dt} \left( \frac{\partial x_i}{\partial q_j} \right) &= \frac{\partial \dot{x_i}}{\partial q_j} \\
   &= \frac{\partial v_i}{\partial q_j} a^{-2}   \\ 
 \frac{d}{dt} \left( \frac{\partial v_i}{\partial q_j} \right) &= \frac{\partial \dot{v_i}}{\partial q_j} \\
 &= - \frac{\partial \phi}{\partial x_i \partial q_j}  a^{-1} \\
 &= - \sum_k  \frac{\partial \phi}{\partial x_i \partial x_k} \frac{\partial x_k}{\partial q_j} a^{-1} \\
 &=: \sum_k T_{ik} D_{xq,kj} a^{-1}
\end{align}
where we defined the tidal tensor $T_{ij} = \partial_i \partial_j \phi$. In matrix form this this can be written as
\begin{align}
  \dot{D}_{xq} &= a^{-2} D_{vq} \\
  \dot{D}_{vq} &= a^{-1} T D_{xq} 
\end{align}
also known as the Geodesic Deviation Equation \citep{vogelsberger_white_2011}. The initial conditions for the distortion tensor are $D_{xq}(a \rightarrow 0) = \mathbb{1}$ and $D_{vq}(a \rightarrow 0) = \mathbb{0}$ in the cosmological case. The maths is worked out in more detail in \cite{vogelsberger_white_2008}.

The stream density of the Lagrangian volume element around a particle can be evaluated as
\begin{align}
  \rho_s &= \frac{\rho_{bg}}{ |\det D_{xq} |}
\end{align}
The Geodesic Deviation Equation allows to follow the exact evolution of an infinitesimal volume element around a particle if the tidal tensor is known. In a simulation the tidal tensor can be evaluated numerically and the GDE can be integrated alongside the other equations of motion \citep{vogelsberger_white_2008, vogelsberger_white_2011}. For a simple analytic model however, the tidal tensor is not known, but we can make certain assumptions about it to get a plausible evolution model.
\subsection{Single-Stream Regions}
In single-stream regions the density $\rho$ is exactly given by the stream density $\rho_s$ of the particles that occupy that point in space. The trace of the tidal tensor therefore has to match
\begin{align}
  \sum_i T_{ii} &= \sum_i \frac{\partial^2 \phi}{\partial x_i \partial x_i} \\
                      &= - \Delta \phi \\
                      &= - 4 \pi G \rho_{bg} \delta \\
                      &= - \frac{4 \pi G}{3} \rho_{bg} \left(\frac{1}{| \det D_{xq} |} -1 \right)
\end{align}
We therefore separate the trace from the trace-free part $T_{ext}$ of the tidal tensor
\begin{align}
  T &= 4 \pi G \rho_{bg} \left(\frac{1}{| \det D_{xq} |} -1 \right) \mathbb{1} + T_{ext}
\end{align}
Note that for a given point in space $T_{ext}$ is completely non-local and depends on the density distribution of the surroundings.
\subsection{Non rotating Tidal Field}
The tidal tensor is symmetric by definition and therefore generally has six components. Since it is symmetric, there is always a system where it is diagonal. In the non linear cosmological density field this system can change its orientation over time leading to all components of the distortion tensor becoming active. However, in linear theory it stays diagonal in the same system at all times. In this case of a tidal field that does not change its orientation we can simplify the Geodesic Deviation Equation, since in the system where the tidal tensor is diagonal also $D_{xq}$ and $D_{vq}$ are diagonal from the beginning, and the non diagonal components never become active. We can then simplify the GDE to six differential equations
\begin{align}
  \dot{x}_i &=   a^{-2} p_i \\
  \dot{p}_i &=   a^{-1} x_i \left( -\frac{4 \pi G}{3} \rho_{bg} \delta + T_{ext, i} \right)
\end{align}
for $i=1,2,3$ where we label by $x_i$ the diagonal components of $D_{xq}$ and by $p_i$ the diagonal components of $D_{vq}$, and use the abbreviation
\begin{align}
  \delta = \frac{1}{| \det D_{xq} |} -1 = \frac{1}{x_1 x_2 x_3} -1
\end{align}
We call models of this form \emph{triaxial collapse models}. They can differ in their assumptions about the external tidal field $T_{ext}$. Note that in the absence of an external tidal field $T_{ext,i} = 0$ this naturally recovers the spherical collapse model.
\subsection{Linear Theory}
In linear theory the tidal tensor takes the form
\begin{align}
    T_{ij} = - 4\pi G \rho_{bg} d_{ij} D(t)
\end{align}
where $d_{ij}$ is the deformation tensor and $D$ is the linear growth factor. $d_{ij}$ is a symmetric tensor and we can transform into the system where it is diagonal with the eigenvalues $d_{ii} = \lambda_i$. The external tidal tensor is then given by
\begin{align}
    T_{ext, ii} &= -4\pi G \rho_{bg} D(t) (\lambda_i - \delta_0/3) \\
    T_{ext, ij} &= 0 \text{   for } i \neq j
\end{align}
where $\delta_0 = \lambda_1 + \lambda_2 + \lambda_3$. We then arrive at
\begin{align}
  \dot{x}_i &=   a^{-2} p_i \\
    \dot{p}_i &= -  \frac{4 \pi G}{3} \rho_{bg} a^{-1} x_i \left( \delta + D(t) (3 \lambda - \delta_0) \right)
\end{align}
where only the external tidal field has been assumed to be given by linear theory but not the local density. Note that it is easy to recover the Zeldovich approximation from this equations by approximating $\delta \approx D(t) \delta_0$, $x_i \sim 1$, leading to
\begin{align}
  \dot{x}_i &=   a^{-2} p_i \\
  \dot{p}_i &= - 4 \pi G \rho_{bg} a^{-1} D(t) \lambda_i
\end{align}
which is solved by $x(t) = 1 + D(t) \lambda$.
\subsection{Fading Tidal Field}
\label{app:tidal}
\begin{figure}
  \includegraphics[width=\columnwidth]{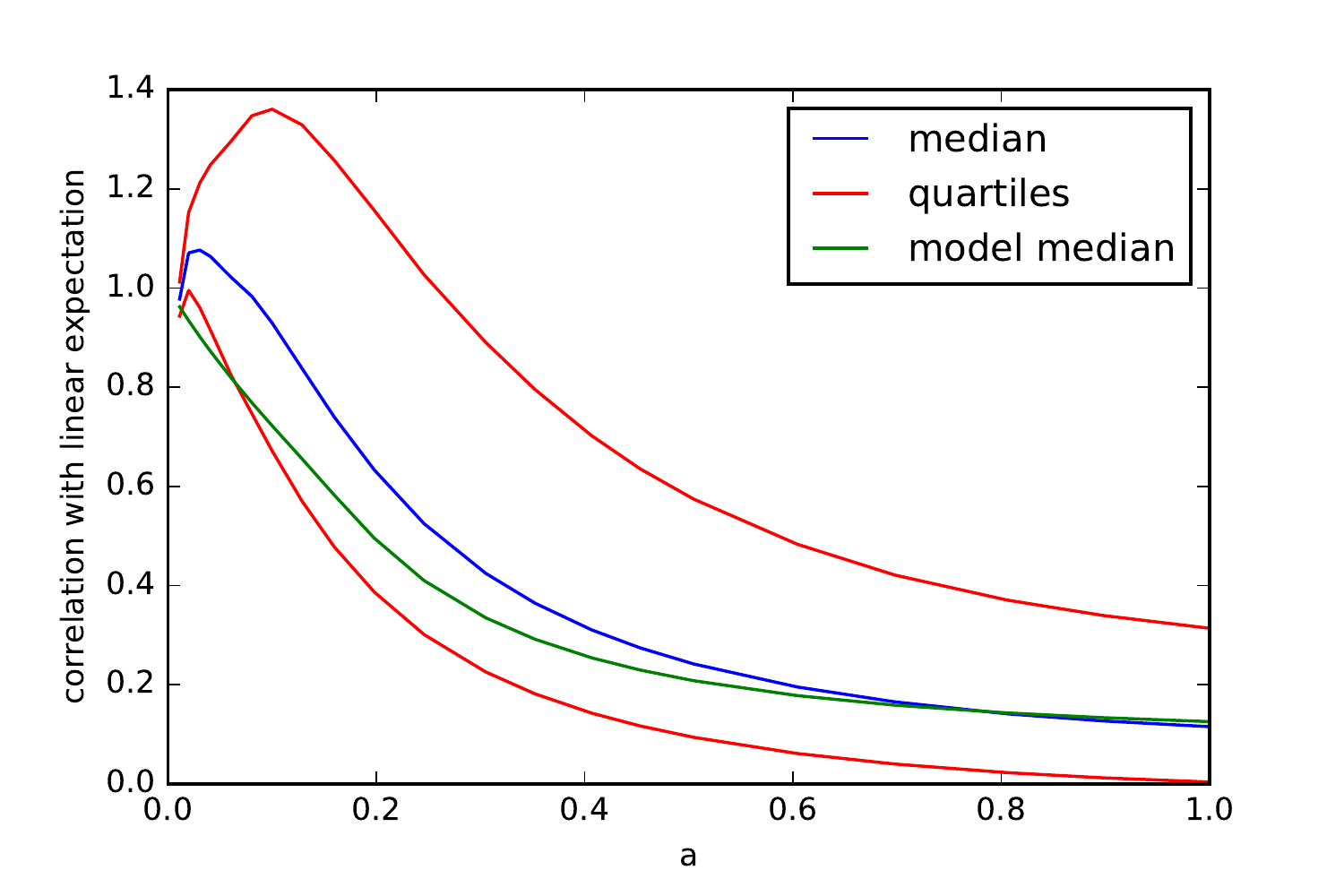}
  \caption{The degree of correspondence of the measured tidal field to the tidal field predicted from linear theory.}
  \label{fig:tidcor}
\end{figure}

While the triaxial collapse model using the external tidal field from linear theory gives a reasonable description for the evolution of volume elements in the linear regime and slightly non linear regime, it certainly cannot be extrapolated to strongly non-linear stages $D(t) \delta_0 \gg 1$. Generally the tidal field from linear theory becomes too large and completely dominates the evolution of volume elements. As can be seen in Figure \ref{fig:tidcor} where we show as a similarity measure between the simulated tidal tensor and the one from linear theory the volume weighted median of the distribution of
\begin{align}
  \beta  =   \frac{\sum_{i,j}T_{ext,sim,ij} T_{ext,lin,ij}}{\sum_{i,j}T_{ext,lin,ij}^2}
\end{align}

In realistic cases however, the non linear tidal field in single-stream regions is becoming smaller and also independent from its initial alignment - therefore becoming sub-dominant in comparison to the local self gravity. Therefore for the excursion set formalism we consider a triaxial collapse model where we multiply  the linear external tidal field by a damping factor to explicitly fade out the tidal forces in the non-linear regime. This leads to an acceleration equation of the form
\begin{align}
\dot{p}_i &= -  \frac{4 \pi G}{3} \rho_{bg} a^{-1} x_i \left( \delta + \alpha(t) (3 \lambda - \delta_0) \right)
\end{align}
where $\alpha$ has to be $D(t)$ in the linear regime and should be limited in the non linear regime. A simple parametrization which seemed to work well for us is
\begin{align}
  \alpha(t) &= \frac{D(t)}{1 + |\delta_0| D(t)} \text{.}
\end{align}
For simplicity we stick to this model for now, but other choices are possible. In principle a more sophisticated model could be found by measuring the statistics of the external tidal field in the non linear regime and quantifying them adequately.

\section{Trigonometric Re-sampling and Single-Stream Regions} \label{app:trigores}
Here we describe an efficient algorithm which can be used to re-sample the dark matter sheet at high resolution using trigonometric interpolation and further explain how it can be modified to distinguish between multi- and single-stream regions in Eulerian space.

\subsection{The Dark Matter Sheet}
Since dark matter is a cold collisonless fluid, it occupies an approximately three-dimensional submanifold of six-dimensional phase-space at all times \citep{white_vogelsberger_2008, abel_2012, shandarin_2012}. This sheet can be reconstructed by tesselation- \citep{abel_2012, shandarin_2012} or interpolation-techniques \citep{hahn_angulo_2016} to get high quality density estimates. We use an interpolation scheme here.

\subsection{Trigonometric Re-sampling}
The particle positions $\vec{x}_i$ of a cosmological simulation can interpreted as sampled values $\vec{x}(\vec{q}_i)$ of the three dimensional vector valued function $\vec{x} (\vec{q})$ where $\vec{q}_i$ are Lagrangian coordinates (i.e. positions in the initial conditions). Since dark matter is a collision-less fluid this function $\vec{x}(\vec{q})$ is continuous and well behaved. 

Therefore interpolation techniques can be used to reconstruct $\vec{x}(\vec{q})$ from the sampling points (particles). For example one can simply order all particles on a grid in Lagrangian space and then interpolate each component $x_i$ of the position $\vec{x}$ individually as a function of the Lagrangian coordinate $\vec{q}$. One can create new re-sampled particles by evaluating this interpolated function at new Lagrangian coordinates $\vec{q}_\text{rs}$. In regions where the interpolation scheme works well, these particles are at the same locations $\vec{x}(\vec{q}_\text{rs})$ as they would have been if one had created an additional particle in the initial conditions at $\vec{q}_\text{rs}$ and followed it through the whole simulation. This interpolation scheme does work well wherever the function $\vec{x}(\vec{q})$ is varying slowly with $\vec{q}$. This is clearly the case in single-stream regions and their boundaries and therefore good enough for our purposes here. Note however, that the interpolation might become poor in the inner parts of haloes, where the function $\vec{x} (\vec{q})$ is varying too quickly to be captured by a limited number of sampling points.

Many different choices are possible and have been used for the interpolation schemes including tesselations \citep{abel_2012, shandarin_2012}, trilinear and triquadratic interpolation \citep{hahn_angulo_2016}. Since we also need a reliable Jacobian here (as we explain in section \ref{app:det_singstr}), we require that all first order derivatives are continuous and therefore we already need at least a third order scheme. Tricubic interpolation would therefore be a viable choice, but unfortunately it is a bit complicated to implement. As an alternative we choose trigonometric interpolation here. Its benefits are that it is simple to implement, it handles the periodic boundary conditions very naturally, all derivatives are well defined, and it is very efficient, since there exist highly optimized algorithms for Fourier transforms.

Therefore we compute the displacement field
\begin{align}
  \vec{s}(\vec{q}) = \vec{x}(\vec{q}) - \vec{q}
\end{align}
which is a periodic function in most cosmological simulations due to the periodic boundary conditions. Note that in practice one has to undo the periodic wrapping of the position function e.g. by wrapping $\vec{s}$ so that all components are in the range $-L/2 \ll s_i \ll L/2$.

Any periodic function on a grid can be simply re-sampled to a higher resolution grid by trigonometric interpolation. Therefore one simply transforms the grid $\vec{s}$ into Fourier space $\vec{s}_k$, zero pads it to a higher resolution grid $\vec{s}_{k, \text{hr}}$, so that all the low frequency components are kept the same and originally unresolved higher frequency components are set to zero, and then transforms the higher resolution Fourier grid back to real-space to obtain the re-sampled displacement field $\vec{s}_{\text{hr}}$.

While this is already an efficient and simple algorithm, it is still a bit impractical in terms of memory consumption, since all re-sampled particles have to be in memory at the same time. Therefore as a more memory efficient version of the same scheme, instead of zero padding the Fourier transformed low resolution grid $\vec{s}_{k}$, we shift it by a small displacement $\Delta \vec{q}$ in Fourier space
\begin{align}
  \vec{s}_{k, \text{shift}} = \exp(i \vec{k} \cdot \Delta \vec{q}) \vec{s}_k
\end{align}
transform it back to real space $\vec{s}_{\text{shift}}$ and then to recover the position field we have to add the shifted Lagrangian coordinate
\begin{align}
  \vec{x}_\text{shift} = \vec{s}_{\text{shift}} + (\vec{q} - \Delta \vec{q})
\end{align}
Now one redoes this shifting for several shifts and each time simply bins the positions $\vec{x}_\text{shift}$ and discards them afterwards. For example to resample from a $2$-dimensional grid with $N^2$ grid cells to a higher resolution grid of $(2N)^2$ grid cells, one applies the 4 shifts 
\begin{align}
 \Delta \vec{q}_{1,2,3,4} = \begin{pmatrix}\pm \Delta g / 4\\ \pm \Delta g / 4 \end{pmatrix} 
\end{align}
where $\Delta g = L/N$ is the gridspacing, $L$ is the boxsize and $N$ the number of grid points per dimension.

\begin{figure}
  \includegraphics[width=\columnwidth]{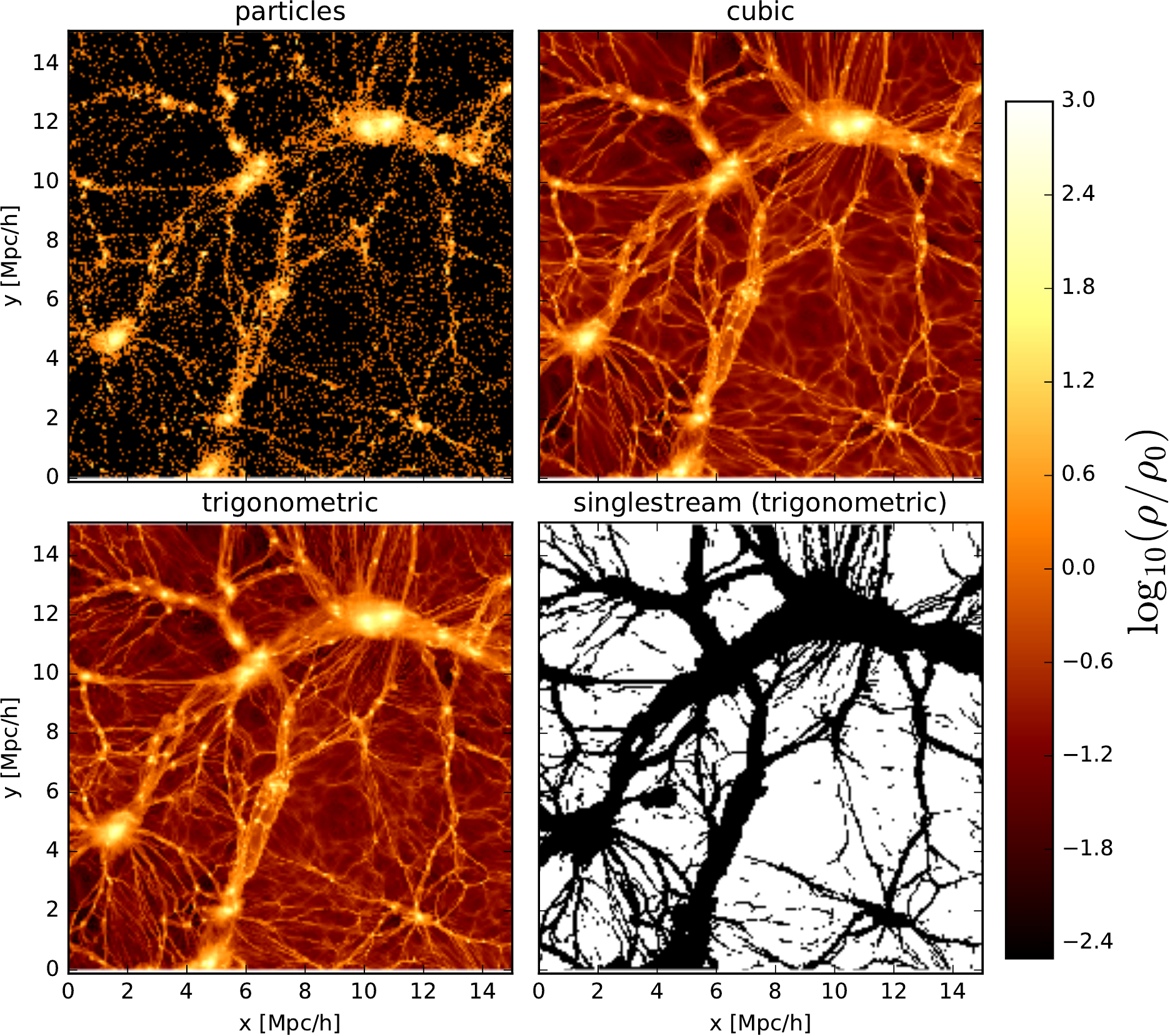}
  \caption{Benchmark of the trigonometric interpolation by projecting a Lagrangian 2D slice from a simulation into Eulerian space. Top Left: density field inferred from binning the particles. Top Right: resampling using third order interpolation. Bottom Left: resampling using trigonometric interpolation. Bottom Right: single-stream regions (white) as inferred by using the trigonometric interpolation. The cubic and trigonometric interpolation agree very well, and the identified single-stream regions match with the ones that one would determine by eye from the density field.}
  \label{fig:trigonometric_interpolation}
\end{figure}

To benchmark the scheme we test it in a two dimensional setup. From a cosmological simulation we select a 2d slice in Lagrangian space (that means all particles that are in one plane in the initial conditions) and project it into Eulerian space with the trigonometric resampling. Further we do the same with a bicubic interpolation as a reference case. Note that - unlike tricubic interpolation - bicubic interpolation is availble as a simple \verb'python' function in \verb'scipy.interpolate.RectBivariateSpline'. We show a zoom into a part of the result in Figure \ref{fig:trigonometric_interpolation}.

\subsection{Determining Single-Stream Regions} \label{app:det_singstr}
The distortion tensor is defined as
\begin{align}
  D :&= \frac{\partial \vec{x}}{\partial \vec{q}} \\
     &= \mathbb{1} + \frac{\partial \vec{s}}{\partial \vec{q}} 
\end{align}
Whenever a particle goes through a caustic the determinant of the distortion tensor flips its sign. All multistream regions contain streams that have gone through one or more caustics, and therefore all multistream regions have a fraction of particles with negative determinant. In contrast, single-stream regions contain exclusively particles with positive determinant. Therefore we determine the distortion tensor for all resampled particles and classify as multistream regions all bins that contain any resampled particles with negative determinant. The distortion tensor of the resampled particles can be simply calculated by differentiating the displacement field in Fourier space:
\begin{align}
  D_{lm} :&= \mathbb{1} + \frac{\partial s_{l} }{\partial q_m} \\
                        &= \mathbb{1} + \text{FT} (i k_m \cdot s_{k, l} )
\end{align}
where FT denotes the (here discrete) Fourier transform.

We show how this method performs on the 2d slice from last section in the bottom right panel of Figure \ref{fig:trigonometric_interpolation}. The determined single-stream regions closely resemble those that one would determine by identifying them by eye from the density field. Note that it is important to resample at high enough resolution to get a continuous field without any gaps. We typically find that resampling at eight times more particles than bins per dimension gives well converged results everywhere.

\label{app:tidal}

\end{document}